\documentclass[11pt]{revtex4}
\usepackage{graphicx}
\usepackage{amssymb}

\makeatletter

\providecommand{\tabularnewline}{\\}

\newcommand{\dblline}[2]{\\ \hline\hline}

\makeatother
\begin{document}

\title{A Stochastic Liouville Equation Approach for the Effect of Noise
in Quantum Computations}

\author{Y. C. Cheng }

\author{R. J. Silbey}

\email{silbey@mit.edu}

\affiliation{Department of Chemistry and Center for Materials Science and Engineering\\
Massachusetts Institute of Technology\\
Cambridge, Massachusetts 02139}

\begin{abstract}
We propose a model based on a generalized effective Hamiltonian for
studying the effect of noise in quantum computations. The system-environment
interactions are taken into account by including stochastic fluctuating
terms in the system Hamiltonian. Treating these fluctuations as Gaussian
Markov processes with zero mean and delta function correlation times,
we derive an exact equation of motion describing the dissipative dynamics
for a system of $n$ qubits. We then apply this model to study the
effect of noise on the quantum teleportation and a generic quantum
controlled-NOT (CNOT) gate. For quantum teleportation, the effect
of noise in the quantum channels are found to be additive, and the
teleportation fidelity depends on the state of the teleported qubit.
The effect of collective decoherence is also studied for the two-qubit
entangled states. For the quantum CNOT gate, we study the effect of
noise on a set of one- and two-qubit quantum gates, and show that
the results can be assembled together to investigate the quality of
a quantum CNOT gate operation. We compute the averaged gate fidelity
and gate purity for the quantum CNOT gate, and investigate phase,
bit-flip, and flip-flop errors during the CNOT gate operation. The
effects of direct inter-qubit coupling and fluctuations on the control
fields are also studied. We find that the quality of the CNOT gate
operation is sensitive to the strengths of the control fields and
the strengths of the noise, and the effect of noise is additive regardless
of its origin. We discuss the limitations and possible extensions
of this model. In sum, we demonstrate a simple model that enables
us to investigate the effect of noise in arbitrary quantum circuits
under realistic device conditions. 
\end{abstract}
\maketitle

\section{INTRODUCTION}

Quantum information processing is of much current interest\cite{nielsen_chuang}.
The realization of quantum algorithms using nuclear magnetic resonance
(NMR) \cite{chuang:nature1998,chuang:prl1998,chuang:prl2000,chuang:nature2001}
and ion-trap \cite{gulde:nature2003} techniques has shown that quantum
computing is possible in principle. Recent efforts for building quantum
computers have focused on techniques based on solid-state devices
that are believed to be more scalable \cite{kane:nature1998,loss:prl1999,makhlin:rmp2001}.
However, such solid-state devices usually require sophisticated manufacturing
techniques, and the inevitable interactions between a qubit and its
surrounding environment ({}``bath'') introduce noise into the quantum
system, resulting in the degradation of the quantum superposition
state. Thus, the extra degrees of freedom of a solid-state system
and the inherent system-bath interactions pose a great problem for
quantum computing with such devices. The decoherence problem is the
main obstacle towards the realization of a universal quantum computer,
and a sound theoretical framework for the description of the decoherence
and population relaxation of qubit systems is necessary~\cite{unruh:pra1995,ekert:pmpe1996}.

Because the ability to compute and predict the behavior of a quantum
circuit under the influence of noise is crucial, a model that can
describe errors from the system-bath interactions could be extremely
useful. Such a model will also be useful in the study of quantum error-correcting
and error-preventing schemes, as well as provide informative guidelines
for the design of quantum computers. However, describing the non-equilibrium
decoherence and population relaxation of a many-qubit system is non-trivial.
No general model exists for this purpose. Classical noise models and
microscopic noise models have yielded some success, but these formulations
do not provide a general solution framework for a many-qubit system.

Classical noise models that describe the decoherence and population
relaxation as exponential decays of the off-diagonal and diagonal
components of the density matrix are widely used for the estimate
of the error rates during quantum computation \cite{knight:pt1997,ekert:pmpe1996},
but generally these models lack quantum features that are important
for quantum computing, such as the quantum interference effect. 

Microscopic noise models based on the spin-boson Hamiltonian that
explicitly include the linear couplings between the system and the
bath degrees of freedom have provided valuable insights about decoherence
effects \cite{walls:pra1985,unruh:pra1995,ekert:pmpe1996}. Recently,
the decoherence and gate performance of a quantum controlled-NOT (CNOT)
gate operation for several different physical realizations has been
studied based on such spin-boson type Hamiltonians \cite{loss:pra1998,hanggi:pra2001,storcz:pra2003,governale:cp2001}.
A number of different techniques has been developed to solve dynamics
of microscopic Hamiltonians \cite{leggett:rmp1987}. However, these
methods are often complicated, and difficult to generalize for systems
with more than two qubits. In addition, in many cases the exact form
of the system-bath interactions is unknown, or the parameters are
difficult to obtain experimentally, and the microscopic models are
difficult to use in these cases.

The Bloch-Redfield formalism is generally used to study NMR spin-dynamics
\cite{argyres:pr1964}, and has been applied to study the dynamics
of many-spin systems \cite{slichter:nmr}. However, this formalism,
while suitable in NMR systems, is not always applicable in general
qubit systems. Moreover, the Bloch-Redfield formalism is also known
to violate the complete positivity of the reduced density operator
at short times. To apply the Bloch-Redfield formalism to quantum computing,
non-physical additional time intervals have to be inserted between
the switching events \cite{loss:pra1998,hanggi:pra2001}. These extra
time periods will result in the over-estimation of the degradation
of the qubit systems. 

Thus, generally speaking, a method that can be used to analyze the
quality of a functional quantum circuit and capable of providing a
quantitative result is still not available. In this work, we propose
a stochastic Liouville equation approach to describe errors in quantum
computations. This approach originates from the Haken-Strobl-Reineker
(HSR) model first proposed by Haken and Strobl and later extended
by Reineker in the 1970s to describe charge and energy transfer in
organic crystals\cite{hr1968,hr1972,reineker:sv1982}. The HSR model
is known to be able to capture the coherent and incoherent dynamics
of quantum two-level systems. In this model, the system-bath interactions
are taken into account by allowing the site energies and the off-diagonal
matrix elements of the system to fluctuate over time. We generalize
the idea of Haken and Strobl to describe a system of $n$ qubits.
The resulting stochastic Liouville equation is then solved to obtain
a set of equations describing the dynamics of a general $n$ qubit
system. To demonstrate the applicability of our method, we study the
effect of noise on quantum teleportation and a generic CNOT gate operation,
and then compare our results with previous work. We show that our
model can reproduce the main results obtained previously by using
microscopic model Hamiltonians. The limitations and possible extensions
of our semiclassical model are also discussed.

\section{THE STOCHASTIC LIOUVILLE EEQUATION APPROACH\label{sec:The-Stochastic-Liouville}}

Previous work on the study of the population relaxation and decoherence
of qubit systems is usually based on the spin-boson Hamiltonian, in
which the qubits are coupled linearly to the bath degrees of freedom
(the environment), and the bath is treated explicitly as a system
of harmonic oscillators \cite{leggett:rmp1987,unruh:pra1995,ekert:pmpe1996}.
Due to the difficulty of applying the spin-boson model to multiple
qubit systems, we take another approach. Instead of treating the bath
explicitly, we follow the stochastic Liouville approach of the HSR
model, and consider an effective Hamiltonian that treats the effect
of the bath as a set of classical fluctuating fields acting on the
system \cite{hr1968,hr1972,reineker:sv1982}. To describe the dynamics
of an array of qubits under the influence of an external control field
and environmental noise, we consider a system of $n$ qubits, and
start from a Hamiltonian with time independent and time dependent
parts. The general Hamiltonian of the qubit system can be written
as

\begin{equation}
\begin{array}{rcl}
\mathbf{H}(t) & = & \mathbf{H}_{0}+\mathbf{h}(t)\\
 & = & {\displaystyle \sum_{i,j=0}^{2^{n}-1}}\left[H_{ij}+h_{ij}(t)\right]c_{i}^{\dagger}c_{j},\end{array}\label{eq:hamiltonian}\end{equation}
where $c_{i}^{\dagger}$ and $c_{i}$ are the creation and annihilation
operators for the $i$-th state of the $2^{n}$ basis set. The time
independent part $\textbf{H}_{0}$ describes the interactions between
qubits, while the time dependent part $\textbf{h}(t)$ describes the
fluctuations of the interactions due to the coupling to the environment.
For simplicity, we assume throughout this work that the external control
fields are switched on and off instantaneously, and the interactions
introduced by the external control fields are constant in time; this
corresponds to a rectangular pulse. More realistic pulse shapes can
be incorporated into our treatment without too much additional work.
In addition, a sequence of different rectangular pulses can be divided
into time periods with a constant external field in each of them,
and then treated separately using a different time independent $\textbf{H}_{0}$
for each time period. Considering only constant external control fields
does not affect the generality of this model.

The time dependent part of the Hamiltonian describes the influences
of the environment via fluctuations of the system energy. This term
may include fluctuations from many different origins, such as the
fluctuations of imperfect control fields, the fluctuations induced
by the bath on the qubit excitation energy, the off-diagonal matrix
element, and the inter-qubit interactions. Following Haken and Strobl
\cite{hr1968}, we consider the fluctuations as random Gaussian Markov
processes with zero mean and $\delta$-function correlation times:

\begin{equation}
\begin{array}{rcl}
\langle h_{ij}(t)\rangle & = & 0\\
\langle h_{ij}(t)h_{kl}(t')\rangle & = & R_{ij;kl}\cdot\delta(t-t')\end{array}\label{eq:stochastic}\end{equation}
Here the brackets $\langle\rangle$ represents the thermal average
over all bath degrees of freedom, and the time independent correlation
matrix element $R_{ij;kl}$ is a real number describing the correlations
between $h_{ij}(t)$ and $h_{kl}(t')$. All $R_{ij;kl}$ elements
form a $2^{2n}$-dimensional correlation matrix $\textbf{R}$. In
addition, we have the following symmetry property of $\textbf{R}$:

\begin{equation}
R_{ij;kl}=R_{ji;kl}=R_{ij;lk}=R_{ji;lk}=R_{kl;ij}\label{eq:symmetry_F}\end{equation}
The value of $R_{ij;kl}$ depends on the strength of the coupling
to the environment; therefore, it is a measure of the noisiness of
the environment. The $\delta$-function correlation in time corresponds
to the limit of fast bath modulations, which assumes that the relaxation
time of the bath is much larger than the characteristic time of the
system. Therefore, this model should be valid at high temperature
limit. Also note that although the effect of the temperature can be
included by considering temperature dependent correlation matrix elements,
there is no explicit temperature dependence in this model. We will
discuss the consequences of this assumption and the applicability
of this model in more detail in Section~\ref{sec:Limitations-of-the}.

The time independent part of the Hamiltonian $\textbf{H}_{0}$ and
the correlation matrix $\textbf{R}$ determine the dynamics of the
system. The values of $\textbf{H}_{0}$ and $\textbf{R}$ depend on
the setup of the physical system, the various types of noise considered,
and the nature of the bath. Note that in the HSR model, the system
is limited to the one exciton subspace, and the matrix $\textbf{H}_{0}$
and $\textbf{R}$ can be obtained directly. However, in our n-qubit
system, all $2^{n}$ states must be considered, and how to determine
$\textbf{H}_{0}$ and $\textbf{R}$ is less obvious. In the following
sections, we provide explicit examples of $\textbf{H}_{0}$ and $\textbf{R}$
for systems describing quantum teleportation and generic quantum gates.
Generalization of the procedure to determine $\textbf{H}_{0}$ and
$\textbf{R}$ for a general n-qubit system should be straightforward.
Throughout this section we will only use the generic forms of $\textbf{H}_{0}$
and $\textbf{R}$ to derive the equation of motion that describes
the time evolution of the $n$-qubit system under the influence of
noise. 

The dynamics of the system is described by the stochastic Liouville
equation ($\hbar=1)$

\[
\dot{\rho}(t)=-i[\mathbf{H}(t),\rho(t)],\]
where $\rho(t)$ is the density matrix of the system at time $t$.
Using the statistical properties of $\textbf{h}(t)$ {[}Eq. (\ref{eq:stochastic}){]}
and the symmetry property of the correlation functions {[}Eq. (\ref{eq:symmetry_F}){]},
we can compute the exact equation of motion for the averaged density
matrix elements of the system by applying the second order generalized
cumulant expansion method to average over all fluctuations. The result
we obtain is in a simple form:

\begin{equation}
\begin{array}{rcl}
\frac{d}{dt}\tilde{\rho}_{\alpha\beta} & = & -i{\displaystyle \sum_{j}H_{\alpha j}\tilde{\rho}_{j\beta}}+i{\displaystyle \sum_{j}\tilde{\rho}_{\alpha j}H_{j\beta}}\\
 &  & -\frac{1}{2}{\displaystyle \sum_{k,l}}R_{lk;k\beta}\tilde{\rho}_{\alpha l}-\frac{1}{2}{\displaystyle \sum_{k,l}}R_{lk;k\alpha}\tilde{\rho}_{l\beta}+{\displaystyle \sum_{k,l}}R_{\beta l;k\alpha}\tilde{\rho}_{kl},\end{array}\label{eq:general_eom}\end{equation}
where all the summations are over all $2^{n}$ state indices. In addition,
we have defined the averaged density matrix of the system, $\tilde{\rho}(t)=\langle\rho(t)\rangle$.
In Eq. (\ref{eq:general_eom}), the dynamics of the averaged density
matrix can be separated into a coherent part, due to $\textbf{H}_{0}$,
and a incoherent part, due to $\textbf{R}$. The dissipation of the
system is governed by incoherent dynamics, and is related to the elements
of the fluctuation correlation matrix $\textbf{R}$. Notice that the
form of Eq. (\ref{eq:general_eom}) is similar to the form of the
widely used Redfield equation, with the relaxation matrix elements
given by the corresponding $R_{ij;kl}$ terms in the equation~\cite{redfield:amr1965}. 

Eq. (\ref{eq:general_eom}) forms a system of $2^{2n}$ linear ordinary
differential equations (ODE). Given the values of $H_{ij}$ and $R_{ij;kl}$,
the ODE system can be solved efficiently to yield the time dependent
averaged density matrix $\tilde{\rho}(t)$. In fact, in most one qubit
and two qubit systems, the equations can be solved analytically, and
the analytical formula for $\tilde{\rho}(t)$ can be obtained. In
general, we can calculate $\textbf{H}_{0}$ and $\textbf{R}$ from
the Hamiltonian of the system and the correlations between fluctuations
introduced by the environment. Once we have $\textbf{H}_{0}$ and
$\textbf{R}$, it is then trivial to solve Eq. (\ref{eq:general_eom})
to yield a $\tilde{\rho}(t)$ that fully describes the dynamics of
the n-qubit system. This procedure is straightforward, and can be
used to study the effect of noise in complex quantum computations.
We demonstrate the applications of this model on the study of the
effect of noise on quantum teleportation and general quantum two qubit
gates in the next two sections.

\section{DISSIPATION IN QUANTUM TELEPORTATION\label{sec:Example-I:-Quantum}}

Since first proposed by Bennett \emph{et al.} in 1993~\cite{bennett:prl1993QT},
the concept of \char`\"{}quantum teleportation\char`\"{} has received
much attention. By exploiting the entangled nature of an Einstein-Podolsky-Rosen
(EPR) pair, a sender can transmit the quantum state of a qubit to
a receiver, without physically transferring the qubit through space.
In this section, we will apply our stochastic Liouville approach to
study the effect of noise on quantum teleportation.

\subsection{Quantum teleportation}

We first consider the ideal scenario of teleporting one qubit from
Alice to Bob. Suppose Alice and Bob share a EPR pair, labeled as qubit
$a$ and $b$, emitted from an EPR pair source, and Alice wants to
teleport qubit $c$ in state $|\psi\rangle=c_{0}|0\rangle+c_{1}|1\rangle$
to Bob. The EPR pair source emits two entangled qubits in one of the
four Bell states at time $t=0$, and then the two qubits are sent
through separate channels $C_{a}$ and $C_{b}$ to Alice and Bob,
respectively. After receiving qubit $a$, Alice performs a Bell-state
measurement on her qubits ($a$ and $c$), and sends the outcome of
her measurement to Bob through a classical channel. Alice's measurement
projects qubit $b$ onto one of the four corresponding states, i.e.
$\mathbf{I}\cdot(c_{0}|0\rangle_{b}+c_{1}|1\rangle_{b})$, $\sigma_{z}\cdot(c_{0}|0\rangle_{b}+c_{1}|1\rangle_{b})$,
$\sigma_{x}\cdot(c_{0}|0\rangle_{b}+c_{1}|1\rangle_{b})$, and $i\sigma_{y}\cdot(c_{0}|0\rangle_{b}+c_{1}|1\rangle_{b})$.
Bob then applies the corresponding inverse transformation ($\mathbf{I}$,
$\sigma_{z}$, $\sigma_{x}$, and $-i\sigma_{y}$, respectively) to
recover his qubit in the state $|\psi\rangle$. 

In practice, errors can happen during the quantum teleportation from
several origins: (1) the noise in the quantum channels $C_{a}$ and
$C_{b}$, (2) the degradation of qubit $c$ after the preparation,
(3) the imperfect Bell-state measurement performed by Alice, (4) the
further degradation of qubit $b$ when transmitting the result of
Bell-state measurement through the classical channel, (5) the imperfect
unitary transformations performed by Bob. Here, we only consider the
first situation where channel $C_{a}$ and $C_{b}$ are noisy, and
focus on the errors due to the degradation of entanglement. We assume
all other operations are done perfectly. This means that result obtained
in the following represents a lower bound on the errors in the quantum
teleportation.

\subsection{Effect of noise on a pair of entangled qubits}

To study the degradation of a pair of entangled qubits, we consider
the effective Hamiltonian describing two uncorrelated qubits $a$
and $b$:

\begin{equation}
\begin{array}{rcl}
\mathbf{H} & = & \mathbf{H_{a}}+\mathbf{H}_{b}\\
 & = & {\displaystyle \sum_{n=a,b}\varepsilon_{n}(t)\cdot\sigma_{z}^{(n)}+\sum_{n=a,b}J_{n}(t)\cdot\sigma_{x}^{(n)}}\\
 & = & {\displaystyle \sum_{n=a,b}[\varepsilon_{n}+\delta\varepsilon_{n}(t)]\cdot\sigma_{z}^{(n)}+\sum_{n=a,b}[J_{n}+\delta J_{n}(t)]\cdot\sigma_{x}^{(n)}},\end{array}\label{eq:H_teleportation}\end{equation}
where $\sigma_{z}^{(n)}$ and $\sigma_{x}^{(n)}$, $n=a,b$ are Pauli
spin operators on qubit $a$ and $b$; $2\varepsilon_{a}$ ($2\varepsilon_{b}$)
is the averaged energy splitting between the $|0\rangle$ and $|1\rangle$
states of qubit $a$ ($b$); $J_{a}$ ($J_{b}$) is the averaged off-diagonal
matrix element for qubit $a$ ($b$); $\delta\varepsilon_{a}(t)$
($\delta\varepsilon_{b}(t)$) is the time-dependent fluctuating part
of the diagonal energy for qubit $a$ ($b$); $\delta J_{a}(t)$ ($\delta J_{b}(t)$)
is the time-dependent fluctuating part of the off-diagonal matrix
element for qubit $a$ ($b$). Following the assumption made in Section
\ref{sec:The-Stochastic-Liouville}, we regard $\delta\varepsilon_{n}(t)$
and $\delta J_{n}(t)$, $n=a,b$ as Gaussian Markov processes fully
described by their first two moments:

\begin{equation}
\begin{array}{rcl}
\langle\delta\varepsilon_{n}(t)\rangle & = & \langle\delta J_{n}(t)\rangle=0,\\
\langle\delta\varepsilon_{n}(t)\delta\varepsilon_{m}(t')\rangle & = & \gamma_{0}^{n}\cdot\delta_{nm}\delta(t-t'),\\
\langle\delta J_{n}(t)\delta J_{m}(t')\rangle & = & \gamma_{1}^{n}\cdot\delta_{nm}\delta(t-t'),\\
\langle\delta\varepsilon_{n}(t)\delta J_{m}(t')\rangle & = & 0,\end{array}\label{eq:teleportation_stochastic}\end{equation}
where $\gamma_{0}^{a}$ ($\gamma_{0}^{b}$ ) describes the strength
of the diagonal energy fluctuations of qubit $a$ ($b$); $\gamma_{1}^{a}$
($\gamma_{1}^{b}$ ) describes the strength of the off-diagonal matrix
element fluctuations of qubit $a$ ($b$). Clearly, $\gamma_{0}^{a}$
and $\gamma_{0}^{b}$ are related to the system-bath interactions
involving $\sigma_{z}$ system operators, and $\gamma_{1}^{a}$ and
$\gamma_{1}^{b}$ are related to the interactions involving $\sigma_{x}$
system operators. These phenomenological parameters can be estimated
experimentally \cite{silbey:arpc1976,reineker:sv1982}. Notice that
we treat the correlation between qubit $a$ and $b$ independently,
because in quantum teleportation, the two EPR qubits are sent through
different channels to two distantly separated places, thus the two
qubits are coupled to distinct baths. In addition, we assume the diagonal
and off-diagonal fluctuations are not correlated. 

To simplify our computations, we choose to study the dynamics of the
system in the Bell-state basis. The four Bell states are defined as

\[
\begin{array}{rcl}
|B_{1}\rangle & = & \frac{1}{\sqrt{2}}(|0\rangle_{a}|0\rangle_{b}+|1\rangle_{a}|1\rangle_{b}),\\
|B_{2}\rangle & = & \frac{1}{\sqrt{2}}(|0\rangle_{a}|0\rangle_{b}-|1\rangle_{a}|1\rangle_{b}),\\
|B_{3}\rangle & = & \frac{1}{\sqrt{2}}(|0\rangle_{a}|1\rangle_{b}+|1\rangle_{a}|0\rangle_{b}),\\
|B_{4}\rangle & = & \frac{1}{\sqrt{2}}(|0\rangle_{a}|1\rangle_{b}-|1\rangle_{a}|0\rangle_{b}),\end{array}\]
where subscript $a,\, b$ labels the state of different qubits. For
convenience, hereafter we will use the notation that use the first
digit to denote the state of qubit $a$, and the second digit to denote
the state of qubit $b$, i.e. $|1\rangle_{a}|1\rangle_{b}\equiv|11\rangle$.
The Hamiltonian for the two qubit system {[}Eq. (\ref{eq:H_teleportation}){]}
in the Bell-state basis is

\begin{equation}
\mathbf{H}=\left[\begin{array}{cccc}
0 & \varepsilon_{a}+\varepsilon_{b}+h_{12}(t) & J_{a}+J_{b}+h_{13}(t) & 0\\
\varepsilon_{a}+\varepsilon_{b}+h_{21}(t) & 0 & 0 & J_{b}-J_{a}+h_{24}(t)\\
J_{a}+J_{b}+h_{31}(t) & 0 & 0 & \varepsilon_{a}-\varepsilon_{b}+h_{34}(t)\\
0 & J_{b}-J_{a}+h_{42}(t) & \varepsilon_{a}-\varepsilon_{b}+h_{43}(t) & 0\end{array}\right],\label{eq:H_tele_transformed}\end{equation}
where the nonzero transformed time-dependent matrix elements are:

\begin{equation}
\begin{array}{rcccl}
h_{12}(t) & = & h_{21}(t) & = & \delta\varepsilon_{a}(t)+\delta\varepsilon_{b}(t),\\
h_{13}(t) & = & h_{31}(t) & = & \delta J_{a}(t)+\delta J_{b}(t),\\
h_{24}(t) & = & h_{42}(t) & = & \delta J_{b}(t)-\delta J_{a}(t),\\
h_{34}(t) & = & h_{43}(t) & = & \delta\varepsilon_{a}(t)-\delta\varepsilon_{b}(t).\end{array}\label{eq:teleportation_transformed_stochastic}\end{equation}

From Eq. (\ref{eq:teleportation_stochastic}) and Eq. (\ref{eq:teleportation_transformed_stochastic}),
we can easily compute the correlation matrix $\textbf{R}$ of the
system. In this case, $\textbf{R}$ has only 32 nonzero elements that
can be represented by the following 6 irreducible elements:

\begin{equation}
\begin{array}{ccc}
R_{12;12} & = & \gamma_{0}^{a}+\gamma_{0}^{b},\\
R_{12;34} & = & \gamma_{0}^{a}-\gamma_{0}^{b},\\
R_{13;13} & = & \gamma_{1}^{a}+\gamma_{1}^{b},\\
R_{13;23} & = & \gamma_{1}^{b}-\gamma_{1}^{a},\\
R_{24;24} & = & \gamma_{1}^{a}+\gamma_{1}^{b},\\
R_{34;34} & = & \gamma_{0}^{a}+\gamma_{0}^{b}.\end{array}\label{eq:teleportation_correlation_matrix}\end{equation}
Other nonzero elements of $\textbf{R}$ can be obtained using the
symmetry property of $\textbf{R}$ (Eq. (\ref{eq:symmetry_F})). Plugging
the correlation matrix elements Eq. (\ref{eq:H_tele_transformed}){]}
and the time-independent Hamiltonian matrix elements {[}Eq. (\ref{eq:teleportation_correlation_matrix}){]}
into Eq. (\ref{eq:general_eom}), we obtain the equation of motion
for the averaged density matrix of the system, $\tilde{\rho}(t)$. 

In the limit of zero averaged Hamiltonian matrix elements, $\varepsilon_{n}=J_{n}=0$,
the equation of motion for the diagonal density matrix elements are
decoupled from those for the off-diagonal density matrix elements.
Therefore, the dynamics of a system initially in one of the four Bell
states (i.e. the initial density matrix has only non-zero diagonal
elements) can be fully described by the equations for the diagonal
density matrix elements:

\begin{equation}
\begin{array}{rcl}
\frac{d}{dt}\tilde{\rho}_{11}(t) & = & \Gamma_{0}\cdot\left[\tilde{\rho}_{22}(t)-\tilde{\rho}_{11}(t)\right]+\Gamma_{1}\cdot\left[\tilde{\rho}_{33}(t)-\tilde{\rho}_{11}(t)\right],\\
\frac{d}{dt}\tilde{\rho}_{22}(t) & = & \Gamma_{0}\cdot\left[\tilde{\rho}_{11}(t)-\tilde{\rho}_{22}(t)\right]+\Gamma_{1}\cdot\left[\tilde{\rho}_{44}(t)-\tilde{\rho}_{22}(t)\right],\\
\frac{d}{dt}\tilde{\rho}_{33}(t) & = & \Gamma_{0}\cdot\left[\tilde{\rho}_{44}(t)-\tilde{\rho}_{33}(t)\right]+\Gamma_{1}\cdot\left[\tilde{\rho}_{11}(t)-\tilde{\rho}_{33}(t)\right],\\
\frac{d}{dt}\tilde{\rho}_{44}(t) & = & \Gamma_{0}\cdot\left[\tilde{\rho}_{33}(t)-\tilde{\rho}_{44}(t)\right]+\Gamma_{1}\cdot\left[\tilde{\rho}_{22}(t)-\tilde{\rho}_{44}(t)\right],\end{array}\label{eq:2qubits_simpleEOM}\end{equation}
where we have defined $\Gamma_{0}=(\gamma_{0}^{a}+\gamma_{0}^{b})$,
and $\Gamma_{1}=(\gamma_{1}^{a}+\gamma_{1}^{b})$. These equations
have the form of a system of kinetic equations involving four states,
and, clearly, $\Gamma_{0}$ and $\Gamma_{1}$ have the meaning of
the degradation rate constants. The symmetric form of Eq. (\ref{eq:2qubits_simpleEOM})
suggests that all four states are equivalent dynamically, hence we
expect the degradation rates of the systems initially in any of the
four Bell states are equal. In this limit, the results of the teleportation
based on different Bell-state channels are the same. Later we will
show that this is only true when $\varepsilon_{n}=J_{n}=0$ and the
two qubits are coupled to distinct baths. 

Eq. (\ref{eq:2qubits_simpleEOM}) also shows that a system of two
qubits initially in one of the maximumly entangled states degrades
into a statistical mixture of the four Bell states. Assuming that
the system is initially in the state $|B_{1}\rangle$ and stays in
the noisy quantum channels for a time period $t$, the density matrix
for the entangled qubits Alice and Bob obtained can be represented
as the statistical mixture \begin{equation}
\tilde{\rho}(t)=\tilde{\rho}_{11}(t)\cdot|B_{1}\rangle\langle B_{1}|+\tilde{\rho}_{22}(t)\cdot|B_{2}\rangle\langle B_{2}|+\tilde{\rho}_{33}(t)\cdot|B_{3}\rangle\langle B_{3}|+\tilde{\rho}_{44}(t)\cdot|B_{4}\rangle\langle B_{4}|.\label{eq:tele_dm_time_t}\end{equation}
The populations can be obtained by solving Eq. (\ref{eq:2qubits_simpleEOM})
with the initial condition $\rho_{0}=|B_{1}\rangle\langle B_{1}|$:

\begin{equation}
\begin{array}{rcl}
\tilde{\rho}_{11}(t) & = & \frac{1}{4}+\frac{1}{4}e^{-2\Gamma_{0}t}+\frac{1}{4}e^{-2\Gamma_{1}t}+\frac{1}{4}e^{-2(\Gamma_{0}+\Gamma_{1})t},\\
\tilde{\rho}_{22}(t) & = & \frac{1}{4}-\frac{1}{4}e^{-2\Gamma_{0}t}+\frac{1}{4}e^{-2\Gamma_{1}t}-\frac{1}{4}e^{-2(\Gamma_{0}+\Gamma_{1})t},\\
\tilde{\rho}_{33}(t) & = & \frac{1}{4}+\frac{1}{4}e^{-2\Gamma_{0}t}-\frac{1}{4}e^{-2\Gamma_{1}t}-\frac{1}{4}e^{-2(\Gamma_{0}+\Gamma_{1})t},\\
\tilde{\rho}_{44}(t) & = & \frac{1}{4}-\frac{1}{4}e^{-2\Gamma_{0}t}-\frac{1}{4}e^{-2\Gamma_{1}t}+\frac{1}{4}e^{-2(\Gamma_{0}+\Gamma_{1})t}.\end{array}\label{eq:simpleEOM_sol}\end{equation}

From Eq. (\ref{eq:simpleEOM_sol}), the fidelity of the entangled
pair, defined as the overlap between the initial density matrix $\rho_{0}$
and the density matrix at time $t$, can be calculated:\begin{equation}
F_{e}(t)=\mathbf{Tr}\rho_{0}\tilde{\rho}(t)=\frac{1}{4}+\frac{1}{4}e^{-2\Gamma_{0}t}+\frac{1}{4}e^{-2\Gamma_{1}t}+\frac{1}{4}e^{-2(\Gamma_{0}+\Gamma_{1})t}.\label{eq:teleportation_fidelity}\end{equation}
Eq. (\ref{eq:teleportation_fidelity}) shows that when $\Gamma_{0}$
and $\Gamma_{1}$ are both non-zero, the fidelity $F_{e}(\infty)=\frac{1}{4}$
in the long time limit. When either $\Gamma_{0}$ or $\Gamma_{1}$
is zero, $F_{e}(\infty)=\frac{1}{2}$. This result indicates that
if we can somehow transform the system and minimize either the diagonal
energy fluctuations or the off-diagonal matrix element fluctuations,
the original quantum state can be better preserved. In addition, Eq.
(\ref{eq:teleportation_fidelity}) can be used to compute a critical
time scale beyond which the degraded entanglement can not be purified
by any entanglement purification method~\cite{bennett:prl1996EP}.
The fidelity required by a successful entanglement purification process,
$F_{e}(t)>0.5$, corresponds to a critical time $t_{c}$ where $F_{e}(t_{c})=\frac{1}{2}$.
For any high-fidelity quantum teleportation to be possible, the EPR
pair should not be allowed to stay in the noisy channels for a time
period longer than $t_{c}$. $t_{c}$ also defines the critical distance
for possible high-fidelity quantum teleportation, given the noise
of the channels described by the parameters $\Gamma_{0}$ and $\Gamma_{1}$.

\subsection{Outcome of teleportation}

Now we can use the result in the previous section to study the outcome
of teleporting a qubit $c$ in state $|\psi\rangle=c_{0}|0\rangle+c_{1}|1\rangle$
from Alice to Bob. We assume the traveling time that the EPR pair
spends in the noisy channels is $t$, and the averaged energy $\varepsilon_{n}$
and off-diagonal matrix elements $J_{n}$ for both qubits are very
small so that the limit of $\varepsilon_{n}=J_{n}=0$, $n=a,b$ can
be applied. After receiving the degraded EPR pair described by Eq.
(\ref{eq:tele_dm_time_t}), Alice and Bob then perform the Bell-state
measurement and corresponding unitary transformation to complete the
teleportation. Assuming that all measurements and unitary transformations
are carried out perfectly and do not introduce more error, the teleportation
outcome that Bob obtains is

\begin{equation}
\rho'(t)=\left[\begin{array}{cc}
\frac{1}{2}+\frac{1}{2}(|c_{0}|^{2}-|c_{1}|^{2})\cdot e^{-2\Gamma_{1}t} & \frac{c_{0}c_{1}^{*}+c_{0}^{*}c_{1}}{2}\cdot e^{-2\Gamma_{0}t}+\frac{c_{0}c_{1}^{*}-c_{0}^{*}c_{1}}{2}\cdot e^{-2(\Gamma_{0}+\Gamma_{1})t}\\
\frac{c_{0}c_{1}^{*}+c_{0}^{*}c_{1}}{2}\cdot e^{-2\Gamma_{0}t}+\frac{c_{0}^{*}c_{1}-c_{0}c_{1}^{*}}{2}\cdot e^{-2(\Gamma_{0}+\Gamma_{1})t} & \frac{1}{2}+\frac{1}{2}(|c_{1}|^{2}-|c_{0}|^{2})\cdot e^{-2\Gamma_{1}t}\end{array}\right].\label{eq:teleportation_outcome}\end{equation}
This result is similar to the result for the dissipation of a two
level system in the HSR model \cite{silbey:arpc1976,reineker:sv1982}.
Notice that the decoherence depends on the total diagonal fluctuations,
$\Gamma_{0}=\gamma_{0}^{a}+\gamma_{0}^{b}$, and the population relaxation
depends on the total off-diagonal fluctuations, $\Gamma_{1}=\gamma_{1}^{a}+\gamma_{1}^{b}$.
Clearly, noise in both channels affect the teleportation outcome additively.
In fact, the outcome is exactly the same as if the teleported qubit
is transfered physically from Alice to Bob through the noisy channel
$C_{a}$ and $C_{b}$, although the qubit Bob receives has never traveled
through channel $C_{a}$ physically. 

The fidelity of teleportation as a function of the traveling time
$t$ is

\begin{equation}
F_{tele}(t)=\frac{1}{2}+\frac{1}{2}(c_{0}^{*}c_{1}+c_{0}c_{1}^{*})^{2}e^{-2\Gamma_{0}t}+\frac{1}{2}(|c_{0}|^{2}-|c_{1}|^{2})^{2}e^{-2\Gamma_{1}t}-\frac{1}{2}(c_{0}^{*}c_{1}-c_{0}c_{1}^{*})^{2}e^{-2(\Gamma_{0}+\Gamma_{1})t}.\label{eq:outcome_fidelity}\end{equation}
The fidelity of teleportation decreases monotonically from $1$ to
$\frac{1}{2}$ as the traveling time $t$ increases. At the long time
limit, the fidelity approaches $\frac{1}{2}$, which means the result
of the quantum teleportation is a half-half mixture of $|0\rangle$
and $|1\rangle$ states, i.e. information about $|\psi\rangle$ is
totally lost. This result is in agreement with recent studies on the
effect of noise on quantum teleportation~\cite{kumar:pra2003}.




Eq. (\ref{eq:outcome_fidelity}) provides a simple interpretation
for the phenomenological parameter $\Gamma_{0}$ and $\Gamma_{1}$:
$\Gamma_{0}$ is the total decay rate for the real part of the coherence,
and $\Gamma_{1}$ is the total population relaxation rate. Recall
that $\gamma_{0}^{n}$, $n=a,b$ is defined using the second moment
of the diagonal energy fluctuation $\delta\varepsilon_{n}(t)$, $n=a,b$
(coupling involving $\sigma_{z}$ ) and $\gamma_{1}^{n}$, $n=a,b$
is defined using the second moment of the off-diagonal matrix element
fluctuation $\delta J_{n}(t)$, $n=a,b$ (coupling involving $\sigma_{x}$
). We see clearly the effects of different types of noise: the diagonal
fluctuations introduce phase shifts that only affect the coherence
of the qubit; the off-diagonal fluctuations introduce coupling between
the two states and result in population transfer. Note that the decay
of the imaginary part of the coherence depends on both diagonal and
off-diagonal fluctuations. In the terminology of quantum computing,
phase-shift errors are caused by the diagonal energy fluctuations,
bit-flip errors are caused by the off-diagonal matrix element fluctuations,
and the change in the $\sigma_{y}$ component are due to both types
of fluctuations. Previous studies on the dissipation of qubits using
spin-boson types of Hamiltonian give similar results for the effects
of different types of system-bath interactions~\cite{walls:pra1985,unruh:pra1995,ekert:pmpe1996}.
Our model gives direct relationships between the phenomenological
parameters describing the strength of the fluctuations and the dissipation
rates. In addition, our model can take into account the effects of
both types of fluctuations \emph{simultaneously}, which is different
from most error models used previously.

\subsection{Nonzero averaged matrix elements}

When the time-independent part of the Hamiltonian contains nonzero
matrix elements, i.e. $\varepsilon_{n}\neq0$ or $J_{n}\neq0$, $n=a,b$,
the exact analytical expression for $\tilde{\rho}(t)$ is not generally
available. In addition, the effect of diagonal energy fluctuations
no longer can be clearly distinguished from the effect of off-diagonal
matrix element fluctuations, both population relaxation and decoherence
depend on $\gamma_{0}^{n}$ and $\gamma_{1}^{n}$, $n=a,b$ . More
importantly, the four Bell states no longer decay at the same rate,
and we can see the effect of the coherent dynamics depending on the
value of the averaged energy and off-diagonal matrix elements. In
the weakly-damped regime where the averaged Hamiltonian matrix elements
are larger than the strength of the noise, the dynamics of a pair
of entangled qubits exhibits coherent oscillating behavior. These
oscillations can lead to errors of the quantum teleportation. Figure
\ref{fig:F_tele_coherent} shows the fidelity of the four Bell states
as a function of traveling time at $\varepsilon_{a}=\varepsilon_{b}=1$,
$J_{a}=J_{b}=0.5$, $\gamma_{0}^{a}=\gamma_{0}^{b}=0.1$, and $\gamma_{1}^{a}=\gamma_{1}^{b}=0.1$.
The different oscillating behavior of the Bell states can be understood
by considering the time-independent part of the Hamiltonian. From
Eq. (\ref{eq:H_teleportation}), all the nonzero time-independent
matrix elements are

\[
\begin{array}{ccccc}
\langle B_{1}|\mathbf{H}_{0}|B_{2}\rangle & = & \langle B_{2}|\mathbf{H}_{0}|B_{1}\rangle & = & \varepsilon_{a}+\varepsilon_{b},\\
\langle B_{1}|\mathbf{H}_{0}|B_{3}\rangle & = & \langle B_{3}|\mathbf{H}_{0}|B_{1}\rangle & = & J_{a}+J_{b},\\
\langle B_{2}|\mathbf{H}_{0}|B_{4}\rangle & = & \langle B_{4}|\mathbf{H}_{0}|B_{2}\rangle & = & J_{b}-J_{a},\\
\langle B_{3}|\mathbf{H}_{0}|B_{4}\rangle & = & \langle B_{4}|\mathbf{H}_{0}|B_{3}\rangle & = & \varepsilon_{a}-\varepsilon_{b}.\end{array}\]
These matrix elements govern the coherent transition between the Bell
states, and result in the oscillating behavior of the dynamics. In
Fig. \ref{fig:F_tele_coherent}, the fidelity of the $|B_{4}\rangle$
state decays monotonically as $t$ increases, because both matrix
elements couple this state to the other states, $J_{b}-J_{a}$ and
$\varepsilon_{a}-\varepsilon_{b}$, are zero for the parameters used.
This also explains why the fidelity of the $|B_{4}\rangle$ state
provides an upper bound on the fidelity of other Bell states in Fig.
\ref{fig:F_tele_coherent}. The state that is coupled most weakly
to other states decay most slowly.

In the regime where the averaged Hamiltonian matrix elements are smaller
than the strength of the noise, the system is overdamped and no oscillating
behavior can be observed. Figure \ref{fig:F_tele_overdamp} shows
the fidelity of the four Bell states at $\varepsilon_{a}=\varepsilon_{b}=0.1$,
$J_{a}=J_{b}=0.05$, $\gamma_{0}^{a}=\gamma_{0}^{b}=0.1$, and $\gamma_{1}^{a}=\gamma_{1}^{b}=0.1$.
In this regime, all Bell states degrade monotonically as the traveling
time increases.

The fidelity of the EPR pair used in the quantum teleportation is
directly related to the fidelity of teleportation. Therefore, the
above discussion can be directly applied to the fidelity of teleportation
performed using different Bell states. When $\varepsilon_{n}\neq0$
or $J_{n}\neq0$, $n=a,b$ , the fidelity of the teleportation behaves
differently when different Bell states are used. To achieve the best
result for the teleportation, we have to choose the Bell state that
is coupled most weakly to other states. In general, $\varepsilon_{n}>0$,
$n=a,b$ and $J_{a}$ and $J_{b}$ have the same sign, thus $|B_{4}\rangle$
state will have the weakest coupling. The singlet $|B_{4}\rangle$
state is the preferred EPR state for the quantum teleportation.\\

\begin{center}{[}Fig. \ref{fig:F_tele_coherent}{]}\\
{[}Fig. \ref{fig:F_tele_overdamp}{]}\end{center}

\subsection{Effect of collective bath}

We have studied the dissipation of two entangled qubits each coupled
to a distinct bath, which is the typical situation relevant for the
quantum teleportation. Another interesting case is when the two qubits
are coupled to a common bath. In this case, we use the Hamiltonian
of Eq. (\ref{eq:H_teleportation}); the difference in the state of
the bath is reflected by different correlation functions for the stochastic
processes. When the two qubits are coupled to a common bath, the first
two moments can be represented as

\begin{equation}
\begin{array}{rcl}
\langle\delta\varepsilon_{n}(t)\rangle & = & \langle\delta J_{n}(t)\rangle=0,\\
\langle\delta\varepsilon_{n}(t)\delta\varepsilon_{m}(t')\rangle & = & \gamma_{0}\cdot\delta(t-t'),\\
\langle\delta J_{n}(t)\delta J_{m}(t')\rangle & = & \gamma_{1}\cdot\delta(t-t'),\\
\langle\delta\varepsilon_{n}(t)\delta J_{m}(t')\rangle & = & 0,\end{array}\label{eq:sambath_teleportation_stochastic}\end{equation}
where $\gamma_{0}$ describes the strength of the diagonal fluctuations;
$\gamma_{1}$ describes the strength of the off-diagonal fluctuations.
Note that because the qubits are coupled to a common bath, the fluctuations
on different qubits are correlated. From Eq. (\ref{eq:teleportation_transformed_stochastic})
and Eq. (\ref{eq:sambath_teleportation_stochastic}), we can derive
the correlation matrix $\textbf{R}$ for the system in the Bell-state
basis. In this collective bath limit, $\textbf{R}$ has only 8 nonzero
elements that can be represented by the following 2 irreducible elements:\begin{equation}
\begin{array}{ccc}
R_{12;12} & = & 4\gamma_{0},\\
R_{13;13} & = & 4\gamma_{1}.\end{array}\label{eq:samebath_teleportation_correlation_matrix}\end{equation}

Using Eq. (\ref{eq:samebath_teleportation_correlation_matrix}), we
can derive the equation of motion for the dynamics of two qubits coupled
to a common bath. In the limit of zero averaged Hamiltonian matrix
elements ($\varepsilon_{n}=J_{n}=0$, $n=a,b$), we obtain a simple
result for the populations in the four Bell states:

\begin{equation}
\begin{array}{rcl}
\frac{d}{dt}\tilde{\rho}_{11}(t) & = & 4\gamma_{0}\cdot\left[\tilde{\rho}_{22}(t)-\tilde{\rho}_{11}(t)\right]+4\gamma_{1}\cdot\left[\tilde{\rho}_{33}(t)-\tilde{\rho}_{11}(t)\right],\\
\frac{d}{dt}\tilde{\rho}_{22}(t) & = & 4\gamma_{0}\cdot\left[\tilde{\rho}_{11}(t)-\tilde{\rho}_{22}(t)\right],\\
\frac{d}{dt}\tilde{\rho}_{33}(t) & = & 4\gamma_{1}\cdot\left[\tilde{\rho}_{11}(t)-\tilde{\rho}_{33}(t)\right],\\
\frac{d}{dt}\tilde{\rho}_{44}(t) & = & 0.\end{array}\label{eq:2q_SameBath_simpleEOM}\end{equation}
Eq. (\ref{eq:2q_SameBath_simpleEOM}) describes the dynamics for a
system of two qubits coupled to a common bath in the Bell-state basis.
Interestingly, the population in the $|B_{4}\rangle$ state, $\tilde{\rho}_{44}(t)$,
is invariant in time. In addition, when only diagonal energy fluctuations
exist ($\gamma_{1}=0$), the population in the $|B_{3}\rangle$ state
is also invariant; when only off-diagonal matrix element fluctuations
exist ($\gamma_{1}=0$), the population in the $|B_{2}\rangle$ state
is invariant. Compared to the result of two qubits coupled to distinct
baths {[}see Eq.(\ref{eq:2qubits_simpleEOM}){]}, Eq. (\ref{eq:2q_SameBath_simpleEOM})
shows that the fluctuations interfere constructively for the $|B_{1}\rangle$
state leading to a faster decay rate, but destructively for the $|B_{4}\rangle$
state. This result can be understood easily in our stochastic model.
In our model, the effect of environment on the system is represented
by a fluctuating field, and the interaction Hamiltonian for the two
qubits is $\mathbf{H}_{int}=\sigma_{i}^{(a)}\cdot V_{a}(t)+\sigma_{i}^{(b)}\cdot V_{b}(t)$
($i=x,z$; $a$ and $b$ are labels for different qubits). When the
two qubits are coupled to a common bath, $V_{a}(t)=V_{b}(t)$, we
can factorize the interaction into the form $\mathbf{H}_{int}=(\sigma_{i}^{(a)}+\sigma_{i}^{(b)})\cdot V_{a}(t)$.
Therefore, any state $|\phi\rangle$ that satisfies $\langle\phi|\sigma_{i}^{(a)}+\sigma_{i}^{(b)}|\phi\rangle=0$
does not interact with the fluctuating field, and is invariant to
the noise. We can see that $\langle B_{3}|\sigma_{z}^{(a)}+\sigma_{z}^{(b)}|B_{3}\rangle=0$
and $\langle B_{4}|\sigma_{z}^{(a)}+\sigma_{z}^{(b)}|B_{4}\rangle=0$,
thus both the $|B_{3}\rangle$ and $|B_{4}\rangle$ states are not
affected by phase-shifting noise; $\langle B_{2}|\sigma_{x}^{(a)}+\sigma_{x}^{(b)}|B_{2}\rangle=0$
and $\langle B_{4}|\sigma_{x}^{(a)}+\sigma_{x}^{(b)}|B_{4}\rangle=0$,
thus both the $|B_{2}\rangle$ and $|B_{4}\rangle$ states are not
affected by bit-flipping noise. This effect of the collective bath
has been verified experimentally \cite{kwiat:science2000}, and studied
in theoretical works related to the ideas of {}``quantum error-avoiding
codes'' \cite{zanardi:prl1997,guo:prl1997} and {}``decoherence-free
subspaces'' \cite{lidar:prl1998_DFS,bacon:pra1999_DFS}. Duan \emph{et
al.} have shown similar result using a Hamiltonian that explicitly
includes the linear coupling terms between the system and the boson
bath \cite{guo:pra1998737,guo:prl1997}. The agreement indicates that
our simple stochastic model can handle both the independent and the
collective bath properly. 

Recently, Kumar and Pandey have studied the effect of noise on quantum
teleportation \cite{kumar:pra2003}. They applied two different models,
a stochastic model and a spin-boson type model, to this problem, and
studied the relative teleportation efficiencies of the Bell states.
Their main result is that for the simple stochastic model, the four
Bell states are equivalent, but for the second model in which the
effect of environment is considered explicitly, the $|B_{4}\rangle$
state is least affected by the noise. We obtain a similar conclusion
using the stochastic Liouville equation approach. Based on our result,
we understand that the $|B_{4}\rangle$ state is the least affected
state because of the assumption of a collective bath, not because
the effect of bath is considered microscopically. Like spin-boson
type models, a simple stochastic model when treated correctly can
provide the same result, and gives a simple picture for the effect
of a collective bath versus a localized bath.

\section{ERRORS IN QUANTUM CNOT GATE\label{sec:Example-II:-A}}

Qubits and quantum gates are the basic elements of quantum computing.
A quantum circuit that performs a particular quantum operation can
be expressed as a composition of elementary quantum gates \cite{divincenzo:pra1995}.
In fact, quantum circuits can be constructed using one- and two-qubit
gates as basic building blocks. For example, the quantum CNOT gate
together with all one-qubit quantum gates form such a set of universal
quantum gates \cite{barenco:pra1995}. In reality, quantum computations
are performed by subjecting an array of qubits under a sequence of
control fields that control the Hamiltonian of the qubit system and
result in specific quantum gate operations. Therefore, we consider
the process of quantum computation as preparing the qubit system in
the initial state, then performing programmed control fields on the
qubits in a sequence of time steps, and finally measuring the output
in the working basis. 

To understand the effect of noise on general quantum computations
and help the implementation of quantum computers, we need a model
that can be used to describe the decoherence and population relaxation
for a system of qubits subjected to external control fields. The decoherence
and gate performance of a CNOT gate on various types of physical realizations
have been studied in Refs. \cite{loss:pra1998,hanggi:pra2001,governale:cp2001,storcz:pra2003,hollenberg:pra2003}.
In particular, Thorwart and Hänggi investigated the decoherence and
dissipation for a generic CNOT gate operation using the numerical
\emph{ab initio} technique of the quasiadiabatic-propagator path integral
(QUAPI). They demonstrated that this numerical method is capable of
describing the full time-resolved dynamics of the two-qubit system
in the presence of noise. To our knowledge, so far, the QUAPI method
is the most sophisticated method that has been applied to study the
decoherence during a CNOT gate operation. In this section, we apply
the stochastic Liouville equation approach to study the same generic
CNOT operation investigated by Thorwart and Hänggi, and show that
our model yields similar results. In general, our model is easier
to extend to many qubit systems than the QUAPI method, and can incorporate
the effects of noise from different sources at the same time.

\subsection{A generic model for 2-qubit quantum gates}

In a physical system, a quantum gate can be expressed by a Hamiltonian
with terms representing the control fields that result in the gate
operation. Consider a elementary step in a quantum gate operation
where the control Hamiltonian is switched on, a generic Hamiltonian
describing the constant external fields and the time-dependent fluctuations
(noise) for a two-qubit system can be written as 

\begin{equation}
\begin{array}{rcl}
\mathbf{H}(t) & = & {\displaystyle \sum_{n=a,b}[\varepsilon_{n}+\delta\varepsilon_{n}(t)]\cdot\sigma_{z}^{(n)}+\sum_{n=a,b}[J_{n}+\delta J_{n}(t)]\cdot\sigma_{x}^{(n)}}\\
 &  & +[g+\delta g(t)]\cdot(\sigma_{+}^{(a)}\sigma_{-}^{(b)}+\sigma_{-}^{(a)}\sigma_{+}^{(b)}),\\
 & \equiv & \mathbf{H}_{0}+\mathbf{h}(t)\end{array}\label{eq:H_generic_gate}\end{equation}
where the two qubits are labeled as qubit $a$ and qubit $b$; the
first two terms comprise the Hamiltonian for two non-interacting qubits
considered in Eq. (\ref{eq:H_teleportation}); the last term represents
the inter-qubit interaction with $\sigma_{\pm}^{(n)}=(\sigma_{x}^{(n)}\mp i\sigma_{y}^{(n)})$,
$n=a,b$; $g$ and $\delta g(t)$ are the time-independent and time-dependent
fluctuating part of the inter-qubit coupling. The controllable fields
are represented by the values of $\varepsilon_{n},$ $J_{n}$, $n=a,b$,
and $g$. Quantum gates can be implemented by switching these fields
on and off in a controlled manner. Notice that the $XY$ type of coupling
is adopted in our model Hamiltonian. This interaction is just an illustrative
example, and does not account for all the possible interactions in
a specific realization of solid-state devices. The real form of the
inter-qubit interaction term depends on the controllable interactions
available for each individual physical implementation. Nevertheless,
our model can handle the other types of interactions as well, and
we expect that the model Hamiltonian we use here can reproduce the
same general physical behavior as other two-qubit Hamiltonians. 

From Eq. (\ref{eq:H_generic_gate}), we can write down the time-independent
part of the Hamiltonian in the standard basis \{$|00\rangle$, $|01\rangle$,
$|10\rangle$,$|11\rangle$\}:

\begin{equation}
\mathbf{H}_{0}=\left[\begin{array}{cccc}
\varepsilon_{a}+\varepsilon_{b} & J_{b} & J_{a} & 0\\
J_{b} & \varepsilon_{a}-\varepsilon_{b} & g & J_{a}\\
J_{a} & g & \varepsilon_{b}-\varepsilon_{a} & J_{b}\\
0 & J_{a} & J_{b} & -\varepsilon_{a}-\varepsilon_{b}\end{array}\right],\label{eq:H_generic_matrix_t-indep}\end{equation}
and the time-dependent part of the Hamiltonian is

\begin{equation}
\mathbf{h}(t)=\left[\begin{array}{cccc}
\delta\varepsilon_{a}(t)+\delta\varepsilon_{b}(t) & \delta J_{b}(t) & \delta J_{a}(t) & 0\\
\delta J_{b}(t) & \delta\varepsilon_{a}(t)-\delta\varepsilon_{b}(t) & \delta g(t) & \delta J_{a}(t)\\
\delta J_{a}(t) & \delta g(t) & \delta\varepsilon_{b}(t)-\delta\varepsilon_{a}(t) & \delta J_{b}(t)\\
0 & \delta J_{a}(t) & \delta J_{b}(t) & -\delta\varepsilon_{a}(t)-\delta\varepsilon_{b}(t)\end{array}\right].\label{eq:H_generic_matrix_t-dep}\end{equation}

Furthermore, we assume the two qubits are close to each other in space,
therefore, we consider the correlation functions suitable for two
qubits coupled to a common bath. Again, we assume the fluctuations
have zero mean and $\delta$-function correlation times. The nonzero
second moments are

\begin{equation}
\begin{array}{rcc}
\langle\delta\varepsilon_{n}(t)\delta\varepsilon_{m}(t')\rangle & = & \gamma_{0}\cdot\delta(t-t'),\\
\langle\delta J_{n}(t)\delta J_{m}(t')\rangle & = & \gamma_{1}\cdot\delta(t-t'),\\
\langle\delta g(t)\delta g(t')\rangle & = & \gamma_{2}\cdot\delta(t-t'),\end{array}\label{eq:sambath_generic_stochastic}\end{equation}
where $\gamma_{0}$ describes the strength of the diagonal energy
fluctuations; $\gamma_{1}$ describes the strength of the off-diagonal
matrix element fluctuations; $\gamma_{2}$ describes the strength
of the fluctuations of the inter-qubit interactions. As we have shown
in the previous section, these phenomenological parameters are related
to the kinetic rate of each separate dissipative process, and can
be easily measured experimentally. Also note that we directly include
the inter-qubit coupling fluctuations, which corresponds to two-qubit
flip-flop errors that are difficult to treat in the microscopic spin-boson
type Hamiltonians. 

Eq. (\ref{eq:sambath_generic_stochastic}) can be used to compute
the elements of the correlation matrix $\textbf{R}$. Using $\textbf{R}$
together with the averaged Hamiltonian matrix elements in Eq. (\ref{eq:H_generic_matrix_t-indep}),
we can obtain the equation of motion describing the dynamics of the
two-qubit system subjected to arbitrary one- and two-qubit control
fields. As a result, we can study the dissipative dynamics of the
qubit system during arbitrary gate operations. Although we only consider
an operation done by a set of constant external fields, the behavior
of more complicated gates that involve more than one step can be studied
by combining the result for each elementary operations. In our model,
the results for a set of universal quantum gates can be assembled
to compute the results for a general quantum circuit.

\subsection{The quantum CNOT gate}

The quantum CNOT gate plays a central role in the quantum computation,
because, as we noted above, the set of all one-qubit gates together
with the CNOT gate is universal \cite{barenco:pra1995}. In the standard
basis \{$|00\rangle$, $|01\rangle$, $|10\rangle$,$|11\rangle$\},
the ideal CNOT gate is represented as

\[
U_{CNOT}^{ideal}=\left[\begin{array}{cccc}
1 & 0 & 0 & 0\\
0 & 1 & 0 & 0\\
0 & 0 & 0 & 1\\
0 & 0 & 1 & 0\end{array}\right].\]
This gate operates on two qubits, and inverts the state of the second
qubit if the first qubit is in the state $|1\rangle$. The CNOT gate
cannot be constructed in one step using our model Hamiltonian. Instead,
we must construct the CNOT gate using multiple elementary one- and
two-qubit gates.

To begin with, we define the following one-qubit rotations on qubit
$a$ and $b$:

\[
U_{nz}(\alpha)=e^{i\frac{\alpha}{2}\sigma_{z}^{(n)}},\,\,\, n=a,b,\]

\[
U_{nx}(\alpha)=e^{i\frac{\alpha}{2}\sigma_{x}^{(n)}},\,\,\, n=a,b,\]
 and the two-qubit operation:

\[
U_{j}(\alpha)=e^{i\alpha(\sigma_{+}^{(a)}\sigma_{-}^{(b)}+\sigma_{-}^{(a)}\sigma_{+}^{(b)})}.\]
All these operations can be easily implemented using our model Hamiltonian
{[}Eq. (\ref{eq:H_generic_gate}){]} (with all control fields set
to zero initially): $U_{nz}(\alpha)$, $n=a,b$, can be done by switching
on $\varepsilon_{n}=-\varepsilon_{0}\cdot\mathrm{Sign(\alpha)}$ for
a time period of $\tau=\frac{\alpha}{2\varepsilon_{0}}$; $U_{nx}(\alpha)$,
$n=a,b$, can be done by switching on $J_{n}=-J_{0}\cdot\mathrm{Sign(\alpha)}$
for a time period of $\tau=\frac{\alpha}{2J_{0}}$; $U_{j}(\alpha)$
can be done by switching on $g=-g_{0}\cdot\mathrm{Sign(\alpha)}$
for a time period of $\tau=\frac{\alpha}{g_{0}}$; where the sign
function $\mathrm{Sign(\alpha)}$ returns $-1$ when $\alpha<0$,
and $1$ when $\alpha>0$. Using the corresponding averaged Hamiltonian
$\textbf{H}_{0}$ for each operations and the correlation matrix presented
in the previous section, the equation of motion describing the dynamics
of the two-qubit system subjected to any of these operations can be
easily obtained. Actually, for arbitrary initial conditions, the analytical
solution for the time-dependent two-qubit density matrix $\rho(t)$
during $U_{nx}(\alpha)$, $U_{nz}(\alpha)$, $n=a,b$, and $U_{j}(\alpha)$
operations are available in the Laplace domain, and can be used to
study arbitrary quantum circuits composed by these three elementary
operations.

The CNOT gate can be expressed by the following sequence of one- and
two-qubit gate operations \cite{makhlin:rmp2001}:

\begin{equation}
U_{CNOT}=U_{bx}(\frac{\pi}{2})U_{bz}(\frac{-\pi}{2})U_{bx}(-\pi)U_{j}(\frac{-\pi}{2})U_{ax}(\frac{-\pi}{2})U_{j}(\frac{\pi}{2})U_{bz}(\frac{-\pi}{2})U_{az}(\frac{-\pi}{2}).\label{eq:cnot_gate_sequence}\end{equation}
 Table (\ref{tab:cnot_gate_sequence}) lists the required control
fields and time span to implement each step using our model Hamiltonian.
In Table (\ref{tab:cnot_gate_sequence}), we use $\varepsilon_{0}$,
$J_{0}$, and $g_{0}$ to denote the strength of the controllable
single-qubit bias, intra-qubit coupling, and inter-qubit $XY$ interaction,
respectively. In addition, we assume that the controllable field strengths
and noise (defined by parameters $\gamma_{0}$, $\gamma_{1}$, and
$\gamma_{2}$ as mentioned in the previous section) for the two qubits
are identical. The value of these parameters should depend on the
specific physical realization of the qubit systems. The total time
required to perform the CNOT gate is $\tau_{cnot}=\pi/2\varepsilon_{0}+\pi/J_{0}+\pi/g_{0}$.
For a typical energy scale of 1 meV (suitable for quantum dot qubits),
the operation time is on the picosecond time scale.\\

\begin{center}{[}Table \ref{tab:cnot_gate_sequence}{]}\\
\end{center}

Using the parameters listed in Table (\ref{tab:cnot_gate_sequence}),
we can calculate the time-dependent two-qubit density matrix $\rho(t)$
during CNOT operations under different noise conditions defined by
$\gamma_{0}$, $\gamma_{1}$, and $\gamma_{2}$. Figure \ref{fig:time_resolv_cnot_11}
shows the time-resolved CNOT operation for two qubits initially in
the $|11\rangle$ state. We set the strengths of the control fields
equal to $1$, i.e. $\varepsilon_{0}=J_{0}=g_{0}=1$. The ideal operation
(solid line) starts at population $1$ in the $|11\rangle$ state,
and ends its total population in the $|10\rangle$ state, showing
a successful CNOT operation. Three different noisy operations are
shown in Fig. \ref{fig:time_resolv_cnot_11}: (1) operation with the
strength of the diagonal energy fluctuations $\gamma_{0}=0.05$ (dashed
line), (2) operation with the strength of the off-diagonal matrix
element fluctuations $\gamma_{1}=0.05$ (dash-dotted line), (3) operation
with the strength of the inter-qubit coupling fluctuations $\gamma_{2}=0.05$
(dotted line). The effect of noise on the CNOT operation can be clearly
seen. In previous work, Thorwart and Hänggi derived the same time-resolved
CNOT operation result \cite{hanggi:pra2001}. Our result is very close
to their numerical \emph{ab initio} QUAPI result. The agreement between
our time-resolved result to the QUAPI result gives us confidence that
our model captures the correct physics.

We use the gate fidelity and gate purity to characterize the performance
of the CNOT gate. Other gate quantifiers including the quantum degree
and entanglement capability are also calculated \cite{zoller:prl1997},
but we do not show the results here because they follow the same trend
as the gate fidelity and gate purity. In our formalism, the density
matrix for the two qubits after the noisy CNOT operation, $\rho(\tau_{cnot})$=$U_{CNOT}\rho_{0}U_{CNOT}^{\dagger}$,
can be calculated for any initial density matrix $\rho_{0}$. Following
Thorwart and Hänggi , we average the gate fidelity and gate purity
over 16 initial states to account for the general performance of the
CNOT gate. The 16 unentangled input states $|\psi_{0}^{ij}\rangle$,
$i,j=1,2,3,4$ are defined as $|\psi_{0}^{ij}\rangle=|\phi_{i}\rangle_{a}\otimes|\phi_{j}\rangle_{b}$
with $|\phi_{1}\rangle=|0\rangle$, $|\phi_{2}\rangle=|1\rangle$,
$|\phi_{3}\rangle=(|0\rangle+|1\rangle)/\sqrt{2}$, $|\phi_{3}\rangle=(|0\rangle+|1\rangle)/\sqrt{2}$,
$|\phi_{4}\rangle=(|0\rangle+i|1\rangle)/\sqrt{2}$, and $a,b$ denoting
the state for different qubits. These states span the Hilbert space
for the two-qubit operations, and should give a reasonable result
for the averaged effect \cite{zoller:prl1997,hanggi:pra2001}.

The gate fidelity is defined as the overlap between the ideal output
and the output of the real gate operation. Using the 16 initial states,
the averaged fidelity can be written as

\[
F=\frac{1}{16}\sum_{i,j=1}^{4}\langle\psi_{out}^{ij}|\rho_{CNOT}^{ij}|\psi_{out}^{ij}\rangle,\]
where we have defined the ideal CNOT output $|\psi_{out}^{ij}\rangle=U_{CNOT}^{ideal}|\psi_{0}^{ij}\rangle$,
and the output of the real CNOT operation $\rho_{CNOT}^{ij}$=$U_{CNOT}|\psi_{0}^{ij}\rangle\langle\psi_{0}^{ij}|U_{CNOT}^{\dagger}$.
The gate fidelity is a measure of how close the real operation is
compared to the ideal operation. For a perfect gate operation, the
gate fidelity should be 1.

Similarly, the averaged gate purity is defined as

\[
P=\frac{1}{16}\sum_{i,j=1}^{4}\mathrm{Tr}((\rho_{CNOT}^{ij})^{2}).\]
The gate purity quantifies the effect of decoherence. For a perfect
gate operation, the gate purity should be 1. \\

\begin{center}{[}Figure \ref{fig:time_resolv_cnot_11}{]}\\
\end{center}

\subsection{Dependence on the noise strength}

The results of the averaged gate fidelity and gate purity as a function
of the strength of each individual type of noise are shown in Fig.
\ref{fig:cnot_F_P}. For our generic study, we again set the strengths
of all the control fields to $1$, i.e. $\varepsilon_{0}=J_{0}=g_{0}=1$.
Clearly, different types of noise cause different amount of errors.
However, they all follow the same trend. The deviations of the gate
fidelity and gate purity from the ideal values, i.e. $1-F$ and $1-P$,
are sensitive to the strength of the noise, and saturate to 0.75 in
the strong noise limit; the value 0.75 corresponds to a fully mixed
state. In the weak noise regime, both $1-F$ and $1-P$ depend linearly
on the noise strength, as expected \cite{hanggi:pra2001,storcz:pra2003}.
The proportionality constant in this case is $\sim10$. In fact, the
proportionality constant depends on the strengths of the control fields,
and reflects the total operation time required to complete the CNOT
gate operation. As the strength of the control field increases, the
total operation time decreases, and the qubits have less time to undergo
the dissipative processes, resulting in less degradation. To minimize
the effect of noise, we need to reduce the proportionality constant,
therefore, we will want to operate the device at the highest control
fields possible. However, the situation will be different if increasing
the strengths of the control fields will also introduce more noise.
We will explicitly discuss the effect of the control-field strength
in the next subsection.

From our results for $\varepsilon_{0}=J_{0}=g_{0}=1$, to achieve
the threshold accuracy of the 0.999~99 level needed for arbitrary
long quantum computations \cite{aharonov:stoc1997,knill:prsla1998,preskill:prsla1998},
one needs to keep the noise strength below the $10^{-6}$ level. Assuming
a characteristic energy scale of 1 meV, this value corresponds to
a decoherence time $\gamma^{-1}$ in the $\mu$s scale, which provides
a serious challenge for experimentalists working on the realization
of solid-state quantum computers.\\

\begin{center}{[}Fig. \ref{fig:cnot_F_P}{]}\\
\end{center}

The linear dependence of $1-F$ and $1-P$ on the noise strengths
also indicates that the effect of \emph{the same} type of noise is
additive in the weak noise regime. To study the additivity of \emph{different}
types of noise, we calculate the averaged CNOT gate fidelity when
different types of noise coexist at the same time. We define the total
error of the CNOT gate operation $E$ as the deviation of the gate
fidelity from the ideal value:

\begin{equation}
E(\gamma_{0},\gamma_{1},\gamma_{2})=1-F(\gamma_{0},\gamma_{1},\gamma_{2}),\label{eq:def_E}\end{equation}
where we have explicitly expressed the total error $E$ as a function
of the three different types of noise strengths:$\gamma_{0}$, $\gamma_{1}$,
and $\gamma_{2}$. In Fig. \ref{fig:cnot_additivity}, we show the
errors of the CNOT gate operation where the different types of noise
coexist, and compare them to the total errors obtained by adding up
the errors caused by the individual type of noise. Clearly, for all
four situations considered, these two lines collapse in the weak noise
regime. The results indicate that errors caused by different types
of noise are additive in the small noise regime. In other words, the
following identity holds in the small noise regime:

\begin{equation}
E(\gamma_{0},\gamma_{1},\gamma_{2})=E(\gamma_{0},0,0)+E(0,\gamma_{1},0)+E(0,0,\gamma_{2}).\label{eq:additive}\end{equation}
Eq. (\ref{eq:additive}) justifies previous studies where different
types of system-bath interactions are treated independently \cite{hanggi:pra2001,storcz:pra2003}.
\\

\begin{center}{[}Fig. \ref{fig:cnot_additivity}{]}\\
\end{center}

\subsection{Dependence on the strength of the inter-qubit coupling}

In the solid-state implementations of qubit devices, the achievable
inter-qubit coupling ($g_{0}$ in our model) is usually weaker than
other single-qubit interactions. Since the time required to finish
a quantum gate operation is inverse proportional to the strength of
the control field used, the weak interaction means long operation
time, hence more errors. The two-qubit gate operations ($U_{j}(\alpha)$
in our model) is usually the bottleneck of quantum computation. In
this section, we analyze the dependence of the quality of the quantum
CNOT gate operation on the strength of the inter-qubit coupling $g_{0}$.

If the strength of the inter-qubit coupling $g_{0}$ can be increased
without introducing any extra disturbance on the system, then we expect
operating the device in the strongest $g_{0}$ achievable will give
the best result. However, physically, applying a stronger field also
means introducing stronger noise due to the imperfectness of the field.
In our model, this means stronger fluctuations on the inter-qubit
$XY$ interaction. The extra noise can be expressed in the value of
the $\gamma_{2}$ term. To incorporate this effect, we allow $\gamma_{2}$
to depend on the strength of the inter-qubit coupling $g_{0}$. Figure
\ref{fig:cnot_g_dependence} shows the errors of the CNOT gate operation
as a function of $g_{0}$ at $\gamma_{0}=0.001$, $\gamma_{1}=0.001$,
$\varepsilon_{0}=1$, and $J_{0}=1$. Three different noise strength
dependences are shown: (1) constant $\gamma_{2}=0.001$ (solid curve),
(2) linear $\gamma_{2}=0.001\cdot(1+g_{0})$ (dashed curve), and (3)
quadratic $\gamma_{2}=0.001\cdot(1+g_{0}^{2})$ (dash-dotted curve).
The three curves show the same behavior in the small $g_{0}$ regime,
in which the operation takes too much time and the system is fully
degraded. As the strength of the coupling $g_{0}$ increases, the
errors decrease due to the shorter operation time. When the strength
of the coupling $g_{0}$ approaches the strengths of other control
fields ($\varepsilon_{0}=J_{0}=1$ in this case), the three curves
start to show different behavior. For both constant and linear $\gamma_{2}$,
the errors generated by other operations ($U_{nz}(\alpha)$ and $U_{nx}(\alpha)$)
dominate the errors of the CNOT gate operation, therefore, increasing
$g_{0}$ gains nothing and the curve saturates. Our result for the
constant $\gamma_{2}$ case is in agreement with the result obtained
previously using the QUAPI method \cite{hanggi:pra2001}. The situation
is different when the strength of the noise depends on $g_{0}$ quadratically.
For this case, the errors start to increase after $g_{0}>1$, because
increasing the inter-qubit coupling $g_{0}$ introduces stronger noise
that cannot be compensated by shorter operation times. Therefore,
in the quadratic case, there exists an optimal $g_{0}$ for the gate
operation. \\

\begin{center}{[}Fig. \ref{fig:cnot_g_dependence}{]}\\
\end{center}

\section{LIMITATIONS AND POSSIBLE EXTENSIONS \label{sec:Limitations-of-the}}

We have shown that the generalized HSR model is flexible for realistic
physical devices. Applications of this model on the effect of noise
on the quantum teleportation and CNOT gate operation give us similar
results compared to previous studies based on microscopic models.
In this section, we will briefly discuss the limitations and possible
extensions of this stochastic Liouville equation approach.

A key step in the HSR model is to replace the microscopic system-bath
interactions by stochastic processes. This procedure has permitted
a full description of the dissipative dynamics of qubit systems and
their response to the external fields. At the same time, we introduce
phenomenological parameters to describe the strengths of fluctuations
($\gamma_{0}$, $\gamma_{1}$, and $\gamma_{2}$ in our model). These
parameters have to be determined experimentally or computed using
a separate microscopic model \cite{hr1972,rackovsky:mp1973,capek:cp1993}.
Generally, $\gamma_{0}$, $\gamma_{1}$, and $\gamma_{2}$ should
depend on temperature and increase as temperature increases. However,
our model lacks explicit temperature dependence for these parameters,
thus cannot be used to study the temperature dependence of the qubit
dynamics. Fortunately, these parameters are directly related to physically
measurable quantities, and can be easily determined by experiments.
In our model, $\gamma_{0}$, $\gamma_{1}$, and $\gamma_{2}$ correspond
to the decoherence rate, population relaxation rate, and inter-qubit
flip-flopping rate, respectively; all of them can be measured by one-
and two-qubit experiments. In addition, recent theoretical studies
on the temperature dependence of the quality of quantum CNOT gate
operation suggest that the temperature dependence of the gate performance
is weak \cite{hanggi:pra2001,storcz:pra2003}, which is reasonable
in the weak coupling regime and the temperature range relevant to
solid-state qubit systems.

The assumption of the fast modulation of the bath might be a more
serious problem for the HSR model. The $\delta$-function correlation
time corresponds to an infinite fast decay of the bath correlations,
which leads to incorrect short time dynamics and long time equilibrium
populations. Palma \emph{et al.} have studied the decoherence of a
qubit and shown that the dynamics exhibits a {}``quiet'' and a {}``quantum''
regime at short times, and a {}``thermal'' regime at long times
\cite{ekert:pmpe1996}. The HSR model assumes that the bath relaxes
infinitely fast, thus neglects the dynamics of the system before bath
relaxation takes place. Although the HSR model cannot predict the
short time dynamics correctly, we expect the physics for longer operations
important for quantum computing are reasonably well captured. The
delta function correlation can be replaced by an exponential function
in time, and the extended model for a dichotomic process has been
solved exactly without further assumptions \cite{reineker:pre1998,reineker:pra1991,palenberg:prb2000,palenberg:jcp2001}.
It will be interesting to apply these extended models to quantum computations
and compare the results with the $\delta$ correlation function results. 

The white noise assumption in the HSR model also corresponds to a
bath with infinite temperature, therefore, the resulting equation
of motion does not satisfy detailed balance at finite temperatures.
As a consequence, the system always relaxes to equal populations regardless
of the energy differences between the states. Extensions of the HSR
model to solve this problem has been proposed in Ref. \cite{capek:cp1993}.
In quantum computing, we are mainly concerned about the dynamics of
an unbiased qubit system, and even when a bias field is applied to
the system to perform gate operations, the time period has to be short
to avoid any population relaxation. Since we will only operate the
quantum computer in the time scale that the population relaxation
is negligible, we expect the violation of the detailed balance condition
will not cause serious problems for applications related to quantum
computing.

The stochastic representation for the dynamics of a quantum two-level
system has been investigated in Refs. \cite{volovich:pra1997} and
\cite{kuzovlev:jetp2003}. The correspondence between the phenomenological
parameters describing the stochastic field ($\gamma_{0}$ and $\gamma_{1}$
in this work) and the two-level system microscopic quantities are
also studied. The stochastic approximation is found to be able to
reproduce the results by Leggett \emph{et al.} for the spin-boson
model \cite{leggett:rmp1987}. Our results presented confirm this
observation. In general, the stochastic Liouville equation approach
presented in this work is applicable in the weak system-bath interaction
limit relevant to quantum computations.

\section{CONCLUSION}

In this work, we present a stochastic Liouville equation approach
that provides a simple way to evaluate the effect of noise in quantum
computations. This approach is generalized from the HSR model. Using
an effective system Hamiltonian that includes the system-bath interactions
as stochastic fluctuating terms with zero mean and delta function
correlation times, we derived the exact equation of motion describing
the dissipative dynamics for a system of $n$ qubits. This generalized
equation of motion is similar to the form of the widely used Redfield
equation, with the relaxation matrix elements given by the corresponding
correlation matrix elements.

We then applied this model to study the dissipative dynamics of a
system of two independent qubits that mimics the EPR pair used in
the quantum teleportation. We showed that the phenomenological parameters
used in our model, i.e. $\gamma_{0}$ and $\gamma_{1}$, correspond
to the decoherence and population relaxation rate, respectively. To
study the effect of noise on quantum teleportation, we calculated
the fidelity of quantum teleportation. We found the effect of noise
in the quantum channels are additive, and the teleportation fidelity
depends on the state of the teleported qubit. When the two EPR qubits
are degenerate and have no intra-qubit coupling, the relative efficiencies
of teleportation for the four Bell states are the same; otherwise,
the singlet state $|B_{4}\rangle$ is the most efficient one. When
the two qubits are coupled to the same bath (collective decoherence
case), the $|B_{1}\rangle$ state is superdecoherent, while the $|B_{4}\rangle$
state is decoherence-free.

Furthermore, we studied a generic two-qubit Hamiltonian containing
$XY$ type inter-qubit interaction. The dissipative dynamics of a
set of one- and two-qubit quantum gates were studied, and the results
were then combined to calculate the averaged gate fidelity and gate
purity for the quantum CNOT gate operation. The dependence of the
quality of the quantum CNOT gate operation on the noise strength and
the strength of the inter-qubit coupling were investigated. We found
that the quality of the CNOT gate operation is sensitive to the noise
strength and the strengths of the control fields. In addition, the
effect of noise is additive regardless of its origin. We compared
our results to Thorwart and Hänggi's results obtained by the numerical
\emph{ab initio} QUAPI technique. In general, our results are in good
agreement with those obtained by the numerical QUAPI method.

We also discussed the limitations of the HSR type approach. The consequences
due to the procedure of replacing the system-bath interactions by
classical fluctuating fields and the assumption of the white noise
were considered, and the possible extensions were noted. Generally,
the application of HSR type model in the weak coupling regime that
is relevant to quantum computing is justified.

Finally, we emphasize that the model presented in this work can be
used to study the dissipative dynamics of a many-qubit system with
direct inter-qubit coupling, imperfectness of the control field, and
other many-qubit effects. The power of this method is in its simple
physical interpretation of the parameters and dissipative dynamics.
In addition, because of the $\delta$-function correlation time assumed
in the model, there is no time-ordering problem for the propagator
computed using Eq. (\ref{eq:general_eom}). The resulting propagator
satisfies complete positivity, therefore no additional time period
has to be inserted between switching events, as will be necessary
for methods based on the Bloch-Redfield formalism. As a result, propagators
computed for simple one- and two-qubit gates can be directly assembled
to study the dissipative dynamics of more complicated quantum circuits.
Since this model can handle noise that affects multiple qubits at
the same time, e.g. the fluctuations in the $XY$ type inter-qubit
interaction that results in flip-flop errors, it will be very interesting
to apply this method to analyze the behavior of a quantum circuit
implementing quantum error-correcting or fault-tolerant codes under
the influence of various multiple-qubit noise. We also expect this
method to be applied to evaluate the quality of quantum circuits under
realistic device conditions. Such theoretical studies will be useful
for the design and implementation of quantum computers.

\begin{acknowledgments}
This work has been partly supported by the National Science Foundation.
\end{acknowledgments}
\bibliographystyle{apsrev}
\bibliography{../qcref,../hsrref}

\begin{thebibliography}{49}
\expandafter\ifx\csname natexlab\endcsname\relax\def\natexlab#1{#1}\fi
\expandafter\ifx\csname bibnamefont\endcsname\relax
  \def\bibnamefont#1{#1}\fi
\expandafter\ifx\csname bibfnamefont\endcsname\relax
  \def\bibfnamefont#1{#1}\fi
\expandafter\ifx\csname citenamefont\endcsname\relax
  \def\citenamefont#1{#1}\fi
\expandafter\ifx\csname url\endcsname\relax
  \def\url#1{\texttt{#1}}\fi
\expandafter\ifx\csname urlprefix\endcsname\relax\def\urlprefix{URL }\fi
\providecommand{\bibinfo}[2]{#2}
\providecommand{\eprint}[2][]{\url{#2}}

\bibitem[{\citenamefont{Nielsen and Chuang}(2000)}]{nielsen_chuang}
\bibinfo{author}{\bibfnamefont{M.}~\bibnamefont{Nielsen}} \bibnamefont{and}
  \bibinfo{author}{\bibfnamefont{I.}~\bibnamefont{Chuang}},
  \emph{\bibinfo{title}{Quantum {C}omputation and {Q}uantum {I}nformation}}
  (\bibinfo{publisher}{Cambridge University Press}, \bibinfo{year}{2000}).

\bibitem[{\citenamefont{Chuang et~al.}(1998{\natexlab{a}})\citenamefont{Chuang,
  Vandersypen, Zhou, Leung, and Lloyd}}]{chuang:nature1998}
\bibinfo{author}{\bibfnamefont{I.}~\bibnamefont{Chuang}},
  \bibinfo{author}{\bibfnamefont{L.}~\bibnamefont{Vandersypen}},
  \bibinfo{author}{\bibfnamefont{X.}~\bibnamefont{Zhou}},
  \bibinfo{author}{\bibfnamefont{D.}~\bibnamefont{Leung}}, \bibnamefont{and}
  \bibinfo{author}{\bibfnamefont{S.}~\bibnamefont{Lloyd}},
  \bibinfo{journal}{Nature} \textbf{\bibinfo{volume}{393}}, \bibinfo{pages}{143
  } (\bibinfo{year}{1998}{\natexlab{a}}).

\bibitem[{\citenamefont{Chuang et~al.}(1998{\natexlab{b}})\citenamefont{Chuang,
  Gershenfeld, and Kubinec}}]{chuang:prl1998}
\bibinfo{author}{\bibfnamefont{I.}~\bibnamefont{Chuang}},
  \bibinfo{author}{\bibfnamefont{N.}~\bibnamefont{Gershenfeld}},
  \bibnamefont{and} \bibinfo{author}{\bibfnamefont{M.}~\bibnamefont{Kubinec}},
  \bibinfo{journal}{Phys. Rev. Lett.} \textbf{\bibinfo{volume}{80}},
  \bibinfo{pages}{3408 } (\bibinfo{year}{1998}{\natexlab{b}}).

\bibitem[{\citenamefont{Vandersypen et~al.}(2000)\citenamefont{Vandersypen,
  Steffen, Breyta, Yannoni, Cleve, and Chuang}}]{chuang:prl2000}
\bibinfo{author}{\bibfnamefont{L.}~\bibnamefont{Vandersypen}},
  \bibinfo{author}{\bibfnamefont{M.}~\bibnamefont{Steffen}},
  \bibinfo{author}{\bibfnamefont{G.}~\bibnamefont{Breyta}},
  \bibinfo{author}{\bibfnamefont{C.}~\bibnamefont{Yannoni}},
  \bibinfo{author}{\bibfnamefont{R.}~\bibnamefont{Cleve}}, \bibnamefont{and}
  \bibinfo{author}{\bibfnamefont{I.}~\bibnamefont{Chuang}},
  \bibinfo{journal}{Phys. Rev. Lett.} \textbf{\bibinfo{volume}{85}},
  \bibinfo{pages}{5452 } (\bibinfo{year}{2000}).

\bibitem[{\citenamefont{Vandersypen et~al.}(2001)\citenamefont{Vandersypen,
  Steffen, Breyta, Yannoni, Sherwood, and Chuang}}]{chuang:nature2001}
\bibinfo{author}{\bibfnamefont{L.}~\bibnamefont{Vandersypen}},
  \bibinfo{author}{\bibfnamefont{M.}~\bibnamefont{Steffen}},
  \bibinfo{author}{\bibfnamefont{G.}~\bibnamefont{Breyta}},
  \bibinfo{author}{\bibfnamefont{C.}~\bibnamefont{Yannoni}},
  \bibinfo{author}{\bibfnamefont{M.}~\bibnamefont{Sherwood}}, \bibnamefont{and}
  \bibinfo{author}{\bibfnamefont{I.}~\bibnamefont{Chuang}},
  \bibinfo{journal}{Nature} \textbf{\bibinfo{volume}{414}}, \bibinfo{pages}{883
  } (\bibinfo{year}{2001}).

\bibitem[{\citenamefont{Gulde et~al.}(2003)\citenamefont{Gulde, Riebe,
  Lancaster, Becher, Eschner, Haffner, Schmidt-Kaler, Chuang, and
  Blatt}}]{gulde:nature2003}
\bibinfo{author}{\bibfnamefont{S.}~\bibnamefont{Gulde}},
  \bibinfo{author}{\bibfnamefont{M.}~\bibnamefont{Riebe}},
  \bibinfo{author}{\bibfnamefont{G.}~\bibnamefont{Lancaster}},
  \bibinfo{author}{\bibfnamefont{C.}~\bibnamefont{Becher}},
  \bibinfo{author}{\bibfnamefont{J.}~\bibnamefont{Eschner}},
  \bibinfo{author}{\bibfnamefont{H.}~\bibnamefont{Haffner}},
  \bibinfo{author}{\bibfnamefont{F.}~\bibnamefont{Schmidt-Kaler}},
  \bibinfo{author}{\bibfnamefont{I.}~\bibnamefont{Chuang}}, \bibnamefont{and}
  \bibinfo{author}{\bibfnamefont{R.}~\bibnamefont{Blatt}},
  \bibinfo{journal}{Nature} \textbf{\bibinfo{volume}{421}}, \bibinfo{pages}{48
  } (\bibinfo{year}{2003}).

\bibitem[{\citenamefont{Kane}(1998)}]{kane:nature1998}
\bibinfo{author}{\bibfnamefont{B.}~\bibnamefont{Kane}},
  \bibinfo{journal}{Nature} \textbf{\bibinfo{volume}{393}}, \bibinfo{pages}{133
  } (\bibinfo{year}{1998}).

\bibitem[{\citenamefont{Imamoglu et~al.}(1999)\citenamefont{Imamoglu,
  Awschalom, Burkard, DiVincenzo, Loss, Sherwin, and Small}}]{loss:prl1999}
\bibinfo{author}{\bibfnamefont{A.}~\bibnamefont{Imamoglu}},
  \bibinfo{author}{\bibfnamefont{D.}~\bibnamefont{Awschalom}},
  \bibinfo{author}{\bibfnamefont{G.}~\bibnamefont{Burkard}},
  \bibinfo{author}{\bibfnamefont{D.}~\bibnamefont{DiVincenzo}},
  \bibinfo{author}{\bibfnamefont{D.}~\bibnamefont{Loss}},
  \bibinfo{author}{\bibfnamefont{M.}~\bibnamefont{Sherwin}}, \bibnamefont{and}
  \bibinfo{author}{\bibfnamefont{A.}~\bibnamefont{Small}},
  \bibinfo{journal}{Phys. Rev. Lett.} \textbf{\bibinfo{volume}{83}},
  \bibinfo{pages}{4204 } (\bibinfo{year}{1999}).

\bibitem[{\citenamefont{Makhlin et~al.}(2001)\citenamefont{Makhlin, Schon, and
  Shnirman}}]{makhlin:rmp2001}
\bibinfo{author}{\bibfnamefont{Y.}~\bibnamefont{Makhlin}},
  \bibinfo{author}{\bibfnamefont{G.}~\bibnamefont{Schon}}, \bibnamefont{and}
  \bibinfo{author}{\bibfnamefont{A.}~\bibnamefont{Shnirman}},
  \bibinfo{journal}{Rev. Mod. Phys.} \textbf{\bibinfo{volume}{73}},
  \bibinfo{pages}{357 } (\bibinfo{year}{2001}).

\bibitem[{\citenamefont{Unruh}(1995)}]{unruh:pra1995}
\bibinfo{author}{\bibfnamefont{W.}~\bibnamefont{Unruh}},
  \bibinfo{journal}{Phys. Rev. A} \textbf{\bibinfo{volume}{51}},
  \bibinfo{pages}{992 } (\bibinfo{year}{1995}).

\bibitem[{\citenamefont{Palma et~al.}(1996)\citenamefont{Palma, Suominen, and
  Ekert}}]{ekert:pmpe1996}
\bibinfo{author}{\bibfnamefont{G.}~\bibnamefont{Palma}},
  \bibinfo{author}{\bibfnamefont{K.-A.} \bibnamefont{Suominen}},
  \bibnamefont{and} \bibinfo{author}{\bibfnamefont{A.}~\bibnamefont{Ekert}},
  \bibinfo{journal}{Proc. R. Soc. Lond. A.} \textbf{\bibinfo{volume}{452}},
  \bibinfo{pages}{567} (\bibinfo{year}{1996}).

\bibitem[{\citenamefont{Knight et~al.}(1997)\citenamefont{Knight, Plenio, and
  Vedral}}]{knight:pt1997}
\bibinfo{author}{\bibfnamefont{P.}~\bibnamefont{Knight}},
  \bibinfo{author}{\bibfnamefont{M.}~\bibnamefont{Plenio}}, \bibnamefont{and}
  \bibinfo{author}{\bibfnamefont{V.}~\bibnamefont{Vedral}},
  \bibinfo{journal}{Philos. Trans. R. Soc. Lond. Ser. A-Math. Phys. Eng. Sci.}
  \textbf{\bibinfo{volume}{355}}, \bibinfo{pages}{2381 }
  (\bibinfo{year}{1997}).

\bibitem[{\citenamefont{Walls and Milburn}(1985)}]{walls:pra1985}
\bibinfo{author}{\bibfnamefont{D.}~\bibnamefont{Walls}} \bibnamefont{and}
  \bibinfo{author}{\bibfnamefont{G.}~\bibnamefont{Milburn}},
  \bibinfo{journal}{Phys. Rev. A} \textbf{\bibinfo{volume}{31}},
  \bibinfo{pages}{2403} (\bibinfo{year}{1985}).

\bibitem[{\citenamefont{Loss and DiVincenzo}(1998)}]{loss:pra1998}
\bibinfo{author}{\bibfnamefont{D.}~\bibnamefont{Loss}} \bibnamefont{and}
  \bibinfo{author}{\bibfnamefont{D.}~\bibnamefont{DiVincenzo}},
  \bibinfo{journal}{Phys. Rev. A} \textbf{\bibinfo{volume}{57}},
  \bibinfo{pages}{120 } (\bibinfo{year}{1998}).

\bibitem[{\citenamefont{Thorwart and H\"{a}nggi}(2002)}]{hanggi:pra2001}
\bibinfo{author}{\bibfnamefont{M.}~\bibnamefont{Thorwart}} \bibnamefont{and}
  \bibinfo{author}{\bibfnamefont{P.}~\bibnamefont{H\"{a}nggi}},
  \bibinfo{journal}{Phys. Rev. A} \textbf{\bibinfo{volume}{65}},
  \bibinfo{pages}{012309} (\bibinfo{year}{2002}).

\bibitem[{\citenamefont{Storcz and Wilhelm}(2003)}]{storcz:pra2003}
\bibinfo{author}{\bibfnamefont{M.}~\bibnamefont{Storcz}} \bibnamefont{and}
  \bibinfo{author}{\bibfnamefont{F.}~\bibnamefont{Wilhelm}},
  \bibinfo{journal}{Phys. Rev. A} \textbf{\bibinfo{volume}{67}},
  \bibinfo{pages}{042319} (\bibinfo{year}{2003}).

\bibitem[{\citenamefont{Governale et~al.}(2001)\citenamefont{Governale,
  Grifoni, and Schon}}]{governale:cp2001}
\bibinfo{author}{\bibfnamefont{M.}~\bibnamefont{Governale}},
  \bibinfo{author}{\bibfnamefont{M.}~\bibnamefont{Grifoni}}, \bibnamefont{and}
  \bibinfo{author}{\bibfnamefont{G.}~\bibnamefont{Schon}},
  \bibinfo{journal}{Chem. Phys.} \textbf{\bibinfo{volume}{268}},
  \bibinfo{pages}{273 } (\bibinfo{year}{2001}).

\bibitem[{\citenamefont{Leggett et~al.}(1987)\citenamefont{Leggett,
  Chakravarty, Dorsey, Fisher, Garg, and Zwerger}}]{leggett:rmp1987}
\bibinfo{author}{\bibfnamefont{A.}~\bibnamefont{Leggett}},
  \bibinfo{author}{\bibfnamefont{S.}~\bibnamefont{Chakravarty}},
  \bibinfo{author}{\bibfnamefont{A.}~\bibnamefont{Dorsey}},
  \bibinfo{author}{\bibfnamefont{M.}~\bibnamefont{Fisher}},
  \bibinfo{author}{\bibfnamefont{A.}~\bibnamefont{Garg}}, \bibnamefont{and}
  \bibinfo{author}{\bibfnamefont{W.}~\bibnamefont{Zwerger}},
  \bibinfo{journal}{Rev. Mod. Phys.} \textbf{\bibinfo{volume}{59}},
  \bibinfo{pages}{1 } (\bibinfo{year}{1987}).

\bibitem[{\citenamefont{Argyres and Kelley}(1964)}]{argyres:pr1964}
\bibinfo{author}{\bibfnamefont{P.}~\bibnamefont{Argyres}} \bibnamefont{and}
  \bibinfo{author}{\bibfnamefont{P.}~\bibnamefont{Kelley}},
  \bibinfo{journal}{Phys. Rev.} \textbf{\bibinfo{volume}{134}},
  \bibinfo{pages}{A98} (\bibinfo{year}{1964}).

\bibitem[{\citenamefont{Slichter}(1996)}]{slichter:nmr}
\bibinfo{author}{\bibfnamefont{C.~P.} \bibnamefont{Slichter}},
  \emph{\bibinfo{title}{Principles of {M}agnetic {R}esonance}}
  (\bibinfo{publisher}{Springer Verlag}, \bibinfo{year}{1996}).

\bibitem[{\citenamefont{Haken and Strobl}(1968)}]{hr1968}
\bibinfo{author}{\bibfnamefont{H.}~\bibnamefont{Haken}} \bibnamefont{and}
  \bibinfo{author}{\bibfnamefont{G.}~\bibnamefont{Strobl}},
  \bibinfo{journal}{Z. Physik} \textbf{\bibinfo{volume}{262}},
  \bibinfo{pages}{135} (\bibinfo{year}{1968}).

\bibitem[{\citenamefont{Haken and Reineker}(1972)}]{hr1972}
\bibinfo{author}{\bibfnamefont{H.}~\bibnamefont{Haken}} \bibnamefont{and}
  \bibinfo{author}{\bibfnamefont{P.}~\bibnamefont{Reineker}},
  \bibinfo{journal}{Z. Physik} \textbf{\bibinfo{volume}{249}},
  \bibinfo{pages}{253} (\bibinfo{year}{1972}).

\bibitem[{\citenamefont{Reineker}(1982)}]{reineker:sv1982}
\bibinfo{author}{\bibfnamefont{P.}~\bibnamefont{Reineker}},
  \emph{\bibinfo{title}{Exciton {D}ynamics in {M}olecular {C}rystals and
  {A}ggregates}} (\bibinfo{publisher}{Springer-Verlag, Berlin},
  \bibinfo{year}{1982}).

\bibitem[{\citenamefont{Redfield}(1965)}]{redfield:amr1965}
\bibinfo{author}{\bibfnamefont{A.}~\bibnamefont{Redfield}},
  \bibinfo{journal}{Adv. Mag. Res.} \textbf{\bibinfo{volume}{1}},
  \bibinfo{pages}{1} (\bibinfo{year}{1965}).

\bibitem[{\citenamefont{Bennett et~al.}(1993)\citenamefont{Bennett, Brassard,
  Crepeau, Jozsa, Peres, and Wootters}}]{bennett:prl1993QT}
\bibinfo{author}{\bibfnamefont{C.}~\bibnamefont{Bennett}},
  \bibinfo{author}{\bibfnamefont{G.}~\bibnamefont{Brassard}},
  \bibinfo{author}{\bibfnamefont{C.}~\bibnamefont{Crepeau}},
  \bibinfo{author}{\bibfnamefont{R.}~\bibnamefont{Jozsa}},
  \bibinfo{author}{\bibfnamefont{A.}~\bibnamefont{Peres}}, \bibnamefont{and}
  \bibinfo{author}{\bibfnamefont{W.}~\bibnamefont{Wootters}},
  \bibinfo{journal}{Phys. Rev. Lett.} \textbf{\bibinfo{volume}{70}},
  \bibinfo{pages}{1895 } (\bibinfo{year}{1993}).

\bibitem[{\citenamefont{Silbey}(1976)}]{silbey:arpc1976}
\bibinfo{author}{\bibfnamefont{R.}~\bibnamefont{Silbey}},
  \bibinfo{journal}{Ann. Rev. Phys. Chem.} \textbf{\bibinfo{volume}{27}},
  \bibinfo{pages}{203} (\bibinfo{year}{1976}).

\bibitem[{\citenamefont{Bennett et~al.}(1996)\citenamefont{Bennett, Brassard,
  Popescu, Schumacher, Smolin, and Wootters}}]{bennett:prl1996EP}
\bibinfo{author}{\bibfnamefont{C.}~\bibnamefont{Bennett}},
  \bibinfo{author}{\bibfnamefont{G.}~\bibnamefont{Brassard}},
  \bibinfo{author}{\bibfnamefont{S.}~\bibnamefont{Popescu}},
  \bibinfo{author}{\bibfnamefont{B.}~\bibnamefont{Schumacher}},
  \bibinfo{author}{\bibfnamefont{J.}~\bibnamefont{Smolin}}, \bibnamefont{and}
  \bibinfo{author}{\bibfnamefont{W.}~\bibnamefont{Wootters}},
  \bibinfo{journal}{Phys. Rev. Lett.} \textbf{\bibinfo{volume}{76}},
  \bibinfo{pages}{722 } (\bibinfo{year}{1996}).

\bibitem[{\citenamefont{Kumar and Pandey}(2003)}]{kumar:pra2003}
\bibinfo{author}{\bibfnamefont{D.}~\bibnamefont{Kumar}} \bibnamefont{and}
  \bibinfo{author}{\bibfnamefont{P.}~\bibnamefont{Pandey}},
  \bibinfo{journal}{Phys. Rev. A} \textbf{\bibinfo{volume}{68}},
  \bibinfo{pages}{012317} (\bibinfo{year}{2003}).

\bibitem[{\citenamefont{Kwiat et~al.}(2000)\citenamefont{Kwiat, Berglund,
  Altepeter, and White}}]{kwiat:science2000}
\bibinfo{author}{\bibfnamefont{P.}~\bibnamefont{Kwiat}},
  \bibinfo{author}{\bibfnamefont{A.}~\bibnamefont{Berglund}},
  \bibinfo{author}{\bibfnamefont{J.}~\bibnamefont{Altepeter}},
  \bibnamefont{and} \bibinfo{author}{\bibfnamefont{A.}~\bibnamefont{White}},
  \bibinfo{journal}{Science} \textbf{\bibinfo{volume}{290}},
  \bibinfo{pages}{498 } (\bibinfo{year}{2000}).

\bibitem[{\citenamefont{Zanardi and Rasetti}(1997)}]{zanardi:prl1997}
\bibinfo{author}{\bibfnamefont{P.}~\bibnamefont{Zanardi}} \bibnamefont{and}
  \bibinfo{author}{\bibfnamefont{M.}~\bibnamefont{Rasetti}},
  \bibinfo{journal}{Phys. Rev. Lett.} \textbf{\bibinfo{volume}{79}},
  \bibinfo{pages}{3306 } (\bibinfo{year}{1997}).

\bibitem[{\citenamefont{Duan and Guo}(1997)}]{guo:prl1997}
\bibinfo{author}{\bibfnamefont{L.}~\bibnamefont{Duan}} \bibnamefont{and}
  \bibinfo{author}{\bibfnamefont{G.}~\bibnamefont{Guo}},
  \bibinfo{journal}{Phys. Rev. Lett.} \textbf{\bibinfo{volume}{79}},
  \bibinfo{pages}{1953 } (\bibinfo{year}{1997}).

\bibitem[{\citenamefont{Lidar et~al.}(1998)\citenamefont{Lidar, Chuang, and
  Whaley}}]{lidar:prl1998_DFS}
\bibinfo{author}{\bibfnamefont{D.}~\bibnamefont{Lidar}},
  \bibinfo{author}{\bibfnamefont{I.}~\bibnamefont{Chuang}}, \bibnamefont{and}
  \bibinfo{author}{\bibfnamefont{K.}~\bibnamefont{Whaley}},
  \bibinfo{journal}{Phys. Rev. Lett.} \textbf{\bibinfo{volume}{81}},
  \bibinfo{pages}{2594 } (\bibinfo{year}{1998}).

\bibitem[{\citenamefont{Bacon et~al.}(1999)\citenamefont{Bacon, Lidar, and
  Whaley}}]{bacon:pra1999_DFS}
\bibinfo{author}{\bibfnamefont{D.}~\bibnamefont{Bacon}},
  \bibinfo{author}{\bibfnamefont{D.}~\bibnamefont{Lidar}}, \bibnamefont{and}
  \bibinfo{author}{\bibfnamefont{K.}~\bibnamefont{Whaley}},
  \bibinfo{journal}{Phys. Rev. A} \textbf{\bibinfo{volume}{60}},
  \bibinfo{pages}{1944 } (\bibinfo{year}{1999}).

\bibitem[{\citenamefont{Duan and Guo}(1998)}]{guo:pra1998737}
\bibinfo{author}{\bibfnamefont{L.}~\bibnamefont{Duan}} \bibnamefont{and}
  \bibinfo{author}{\bibfnamefont{G.}~\bibnamefont{Guo}},
  \bibinfo{journal}{Phys. Rev. A} \textbf{\bibinfo{volume}{57}},
  \bibinfo{pages}{737 } (\bibinfo{year}{1998}).

\bibitem[{\citenamefont{DiVincenzo}(1995)}]{divincenzo:pra1995}
\bibinfo{author}{\bibfnamefont{D.}~\bibnamefont{DiVincenzo}},
  \bibinfo{journal}{Phys. Rev. A} \textbf{\bibinfo{volume}{51}},
  \bibinfo{pages}{1015 } (\bibinfo{year}{1995}).

\bibitem[{\citenamefont{Barenco et~al.}(1995)\citenamefont{Barenco, Bennett,
  Cleve, DiVincenzo, Margolus, Shor, Sleator, Smolin, and
  Weinfurter}}]{barenco:pra1995}
\bibinfo{author}{\bibfnamefont{A.}~\bibnamefont{Barenco}},
  \bibinfo{author}{\bibfnamefont{C.}~\bibnamefont{Bennett}},
  \bibinfo{author}{\bibfnamefont{R.}~\bibnamefont{Cleve}},
  \bibinfo{author}{\bibfnamefont{D.}~\bibnamefont{DiVincenzo}},
  \bibinfo{author}{\bibfnamefont{N.}~\bibnamefont{Margolus}},
  \bibinfo{author}{\bibfnamefont{P.}~\bibnamefont{Shor}},
  \bibinfo{author}{\bibfnamefont{T.}~\bibnamefont{Sleator}},
  \bibinfo{author}{\bibfnamefont{J.}~\bibnamefont{Smolin}}, \bibnamefont{and}
  \bibinfo{author}{\bibfnamefont{H.}~\bibnamefont{Weinfurter}},
  \bibinfo{journal}{Phys. Rev. A} \textbf{\bibinfo{volume}{52}},
  \bibinfo{pages}{3457 } (\bibinfo{year}{1995}).

\bibitem[{\citenamefont{Fowler et~al.}(2003)\citenamefont{Fowler, Wellard, and
  Hollenberg}}]{hollenberg:pra2003}
\bibinfo{author}{\bibfnamefont{A.}~\bibnamefont{Fowler}},
  \bibinfo{author}{\bibfnamefont{C.}~\bibnamefont{Wellard}}, \bibnamefont{and}
  \bibinfo{author}{\bibfnamefont{L.}~\bibnamefont{Hollenberg}},
  \bibinfo{journal}{Phys. Rev. A} \textbf{\bibinfo{volume}{67}},
  \bibinfo{pages}{012301} (\bibinfo{year}{2003}).

\bibitem[{\citenamefont{Poyatos et~al.}(1997)\citenamefont{Poyatos, Cirac, and
  Zoller}}]{zoller:prl1997}
\bibinfo{author}{\bibfnamefont{J.}~\bibnamefont{Poyatos}},
  \bibinfo{author}{\bibfnamefont{J.}~\bibnamefont{Cirac}}, \bibnamefont{and}
  \bibinfo{author}{\bibfnamefont{P.}~\bibnamefont{Zoller}},
  \bibinfo{journal}{Phys. Rev. Lett.} \textbf{\bibinfo{volume}{78}},
  \bibinfo{pages}{390 } (\bibinfo{year}{1997}).

\bibitem[{\citenamefont{Aharonov and Ben-Or}(1997)}]{aharonov:stoc1997}
\bibinfo{author}{\bibfnamefont{D.}~\bibnamefont{Aharonov}} \bibnamefont{and}
  \bibinfo{author}{\bibfnamefont{M.}~\bibnamefont{Ben-Or}}, in
  \emph{\bibinfo{booktitle}{Proceedings of the 29 {A}nnual {ACM} {S}ymposium on
  the {T}heory of {C}omputing}} (\bibinfo{year}{1997}), pp.
  \bibinfo{pages}{176--188}.

\bibitem[{\citenamefont{Knill et~al.}(1998)\citenamefont{Knill, Laflamme, and
  Zurek}}]{knill:prsla1998}
\bibinfo{author}{\bibfnamefont{E.}~\bibnamefont{Knill}},
  \bibinfo{author}{\bibfnamefont{R.}~\bibnamefont{Laflamme}}, \bibnamefont{and}
  \bibinfo{author}{\bibfnamefont{W.}~\bibnamefont{Zurek}},
  \bibinfo{journal}{Proc. R. Soc. London A} \textbf{\bibinfo{volume}{454}},
  \bibinfo{pages}{365 } (\bibinfo{year}{1998}).

\bibitem[{\citenamefont{Preskill}(1998)}]{preskill:prsla1998}
\bibinfo{author}{\bibfnamefont{J.}~\bibnamefont{Preskill}},
  \bibinfo{journal}{Proc. R. Soc. London A} \textbf{\bibinfo{volume}{454}},
  \bibinfo{pages}{385 } (\bibinfo{year}{1998}).

\bibitem[{\citenamefont{Rackovsky and Silbey}(1973)}]{rackovsky:mp1973}
\bibinfo{author}{\bibfnamefont{S.}~\bibnamefont{Rackovsky}} \bibnamefont{and}
  \bibinfo{author}{\bibfnamefont{R.}~\bibnamefont{Silbey}},
  \bibinfo{journal}{Mol. Phys.} \textbf{\bibinfo{volume}{25}},
  \bibinfo{pages}{61} (\bibinfo{year}{1973}).

\bibitem[{\citenamefont{\v{C}\'{a}pek}(1993)}]{capek:cp1993}
\bibinfo{author}{\bibfnamefont{V.}~\bibnamefont{\v{C}\'{a}pek}},
  \bibinfo{journal}{Chem. Phys.} \textbf{\bibinfo{volume}{171}},
  \bibinfo{pages}{79 } (\bibinfo{year}{1993}).

\bibitem[{\citenamefont{Warns et~al.}(1998)\citenamefont{Warns, Barvik, and
  Reineker}}]{reineker:pre1998}
\bibinfo{author}{\bibfnamefont{C.}~\bibnamefont{Warns}},
  \bibinfo{author}{\bibfnamefont{I.}~\bibnamefont{Barvik}}, \bibnamefont{and}
  \bibinfo{author}{\bibfnamefont{P.}~\bibnamefont{Reineker}},
  \bibinfo{journal}{Phys. Rev. E} \textbf{\bibinfo{volume}{57}},
  \bibinfo{pages}{3928 } (\bibinfo{year}{1998}).

\bibitem[{\citenamefont{Kraus and Reineker}(1991)}]{reineker:pra1991}
\bibinfo{author}{\bibfnamefont{V.}~\bibnamefont{Kraus}} \bibnamefont{and}
  \bibinfo{author}{\bibfnamefont{P.}~\bibnamefont{Reineker}},
  \bibinfo{journal}{Phys. Rev. A} \textbf{\bibinfo{volume}{43}},
  \bibinfo{pages}{4182 } (\bibinfo{year}{1991}).

\bibitem[{\citenamefont{Palenberg et~al.}(2000)\citenamefont{Palenberg, Silbey,
  and Pfluegl}}]{palenberg:prb2000}
\bibinfo{author}{\bibfnamefont{M.}~\bibnamefont{Palenberg}},
  \bibinfo{author}{\bibfnamefont{R.}~\bibnamefont{Silbey}}, \bibnamefont{and}
  \bibinfo{author}{\bibfnamefont{W.}~\bibnamefont{Pfluegl}},
  \bibinfo{journal}{Phys. Rev. B} \textbf{\bibinfo{volume}{62}},
  \bibinfo{pages}{3744 } (\bibinfo{year}{2000}).

\bibitem[{\citenamefont{Palenberg et~al.}(2001)\citenamefont{Palenberg, Silbey,
  Warns, and Reineker}}]{palenberg:jcp2001}
\bibinfo{author}{\bibfnamefont{M.}~\bibnamefont{Palenberg}},
  \bibinfo{author}{\bibfnamefont{R.}~\bibnamefont{Silbey}},
  \bibinfo{author}{\bibfnamefont{C.}~\bibnamefont{Warns}}, \bibnamefont{and}
  \bibinfo{author}{\bibfnamefont{P.}~\bibnamefont{Reineker}},
  \bibinfo{journal}{J. Chem. Phys.} \textbf{\bibinfo{volume}{114}},
  \bibinfo{pages}{4386 } (\bibinfo{year}{2001}).

\bibitem[{\citenamefont{Accardi et~al.}(1997)\citenamefont{Accardi, Kozyrev,
  and Volovich}}]{volovich:pra1997}
\bibinfo{author}{\bibfnamefont{L.}~\bibnamefont{Accardi}},
  \bibinfo{author}{\bibfnamefont{S.}~\bibnamefont{Kozyrev}}, \bibnamefont{and}
  \bibinfo{author}{\bibfnamefont{I.}~\bibnamefont{Volovich}},
  \bibinfo{journal}{Phys. Rev. A} \textbf{\bibinfo{volume}{56}},
  \bibinfo{pages}{2557 } (\bibinfo{year}{1997}).

\bibitem[{\citenamefont{Kuzovlev}(2003)}]{kuzovlev:jetp2003}
\bibinfo{author}{\bibfnamefont{Y.}~\bibnamefont{Kuzovlev}},
  \bibinfo{journal}{JETP Lett.} \textbf{\bibinfo{volume}{78}},
  \bibinfo{pages}{92 } (\bibinfo{year}{2003}).

\end{thebibliography}

\clearpage \newpage

\begin{table}

\caption{\label{tab:cnot_gate_sequence} Parameters of the model Hamiltonians
used to perform the CNOT gate in $7$ steps. The required control
fields and time span for each step are listed. Note that we only list
the nonzero field parameters.}

\begin{center}\begin{tabular}{cccc}
\hline 
\hline No.&
Operation&
Control Fields&
Time\tabularnewline
\hline
1&
~~~$U_{bz}(\frac{-\pi}{2})U_{az}(\frac{-\pi}{2})$~~~&
~~~$\varepsilon_{a}=\varepsilon_{0}$, $\varepsilon_{b}=\varepsilon_{0}$~~~&
$\tau_{1}=\frac{\pi}{4\varepsilon_{0}}$\tabularnewline
2&
$U_{j}(\frac{\pi}{2})$&
$g=-g_{0}$&
$\tau_{2}=\tau_{1}+\frac{\pi}{2g_{0}}$\tabularnewline
3&
$U_{ax}(\frac{-\pi}{2})$&
$J_{a}=J_{0}$&
$\tau_{3}=\tau_{2}+\frac{\pi}{4J_{0}}$\tabularnewline
4&
$U_{j}(\frac{-\pi}{2})$&
$g=g_{0}$&
$\tau_{4}=\tau_{3}+\frac{\pi}{2g_{0}}$\tabularnewline
5&
$U_{bx}(-\pi)$&
$J_{b}=J_{0}$&
$\tau_{5}=\tau_{4}+\frac{\pi}{2J_{0}}$\tabularnewline
6&
$U_{bz}(\frac{-\pi}{2})$&
$\varepsilon_{b}=\varepsilon_{0}$&
$\tau_{6}=\tau_{5}+\frac{\pi}{4\varepsilon_{0}}$\tabularnewline
7&
$U_{bx}(\frac{\pi}{2})$&
$J_{b}=-J_{0}$&
$\tau_{7}=\tau_{6}+\frac{\pi}{4J_{0}}$\dblline\tabularnewline
\hline
\end{tabular}\end{center}
\end{table}

\clearpage \newpage

\begin{figure}
\begin{center}\includegraphics[%
  width=1.0\columnwidth,
  keepaspectratio]{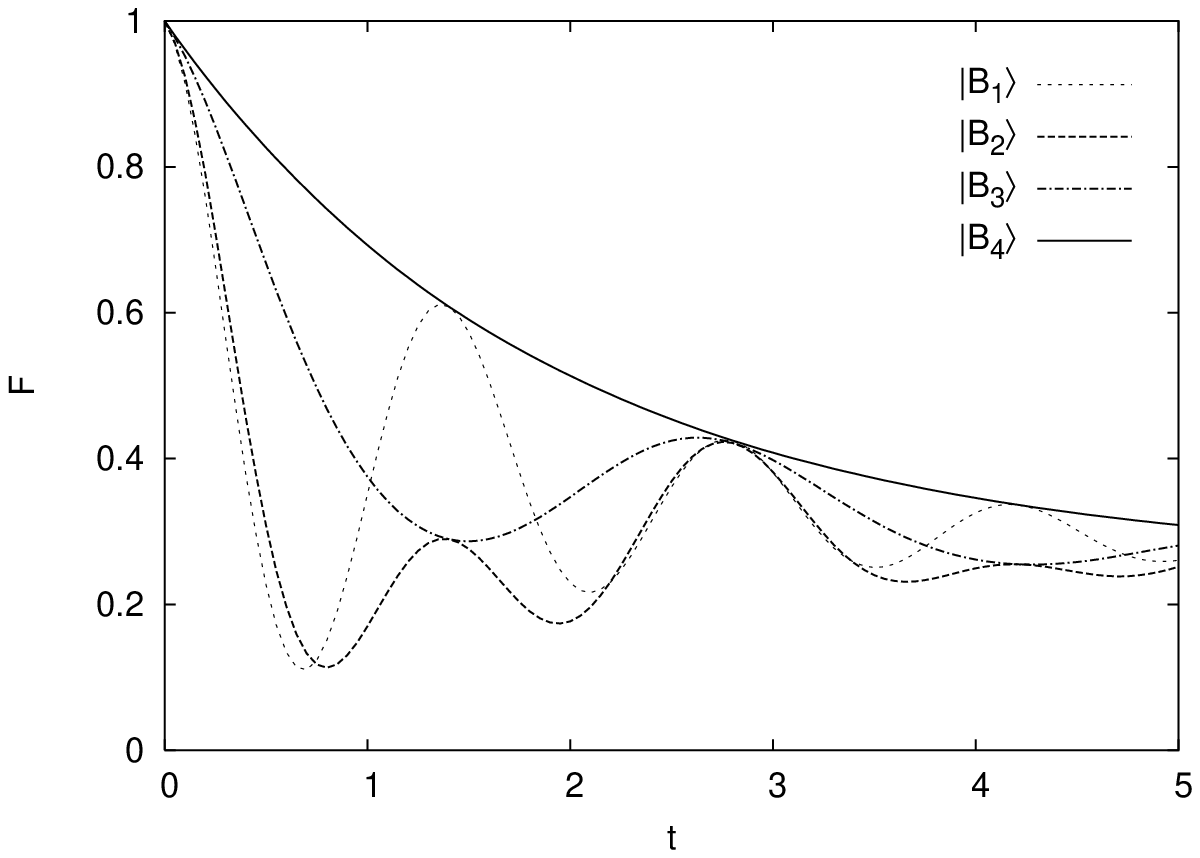}\end{center}

\caption{\label{fig:F_tele_coherent} }
\end{figure}

\begin{figure}
\begin{center}\includegraphics[%
  width=1.0\columnwidth,
  keepaspectratio]{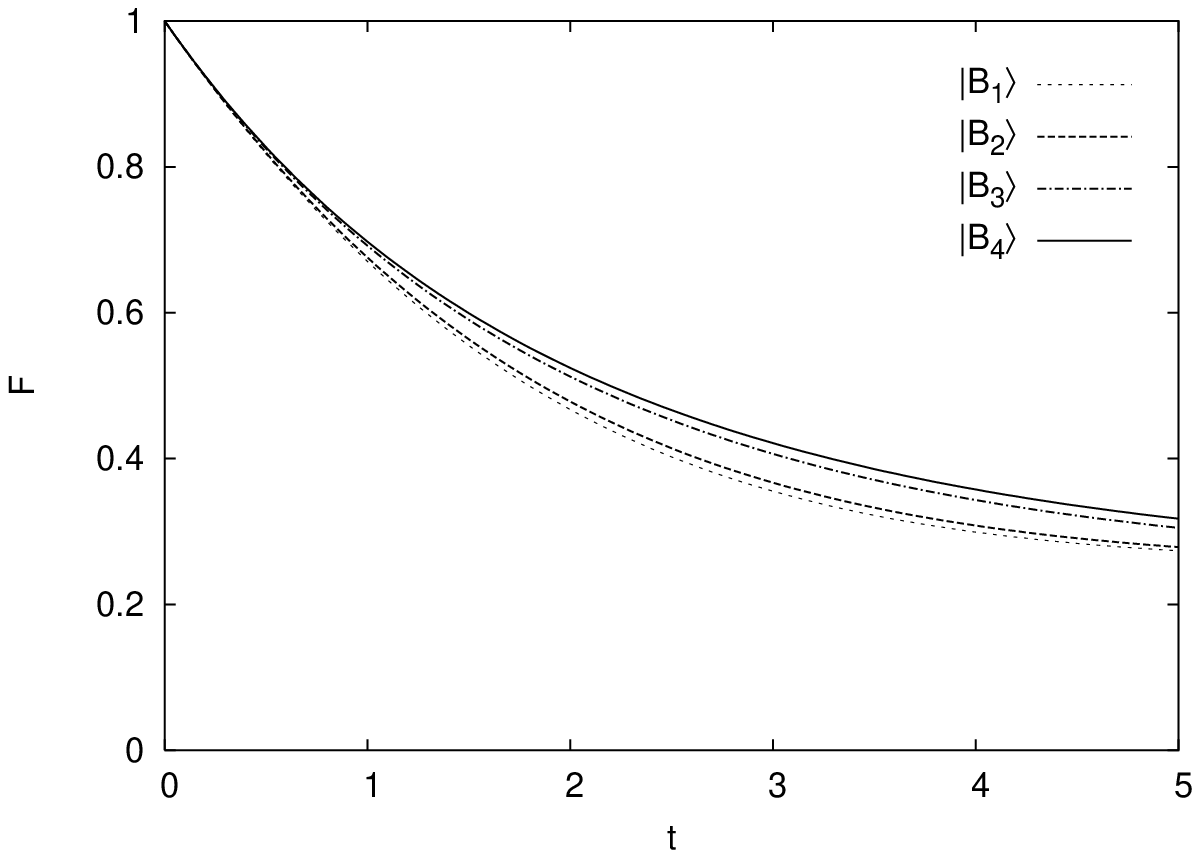}\end{center}

\caption{\label{fig:F_tele_overdamp} }
\end{figure}

\begin{figure}
\begin{center}\includegraphics{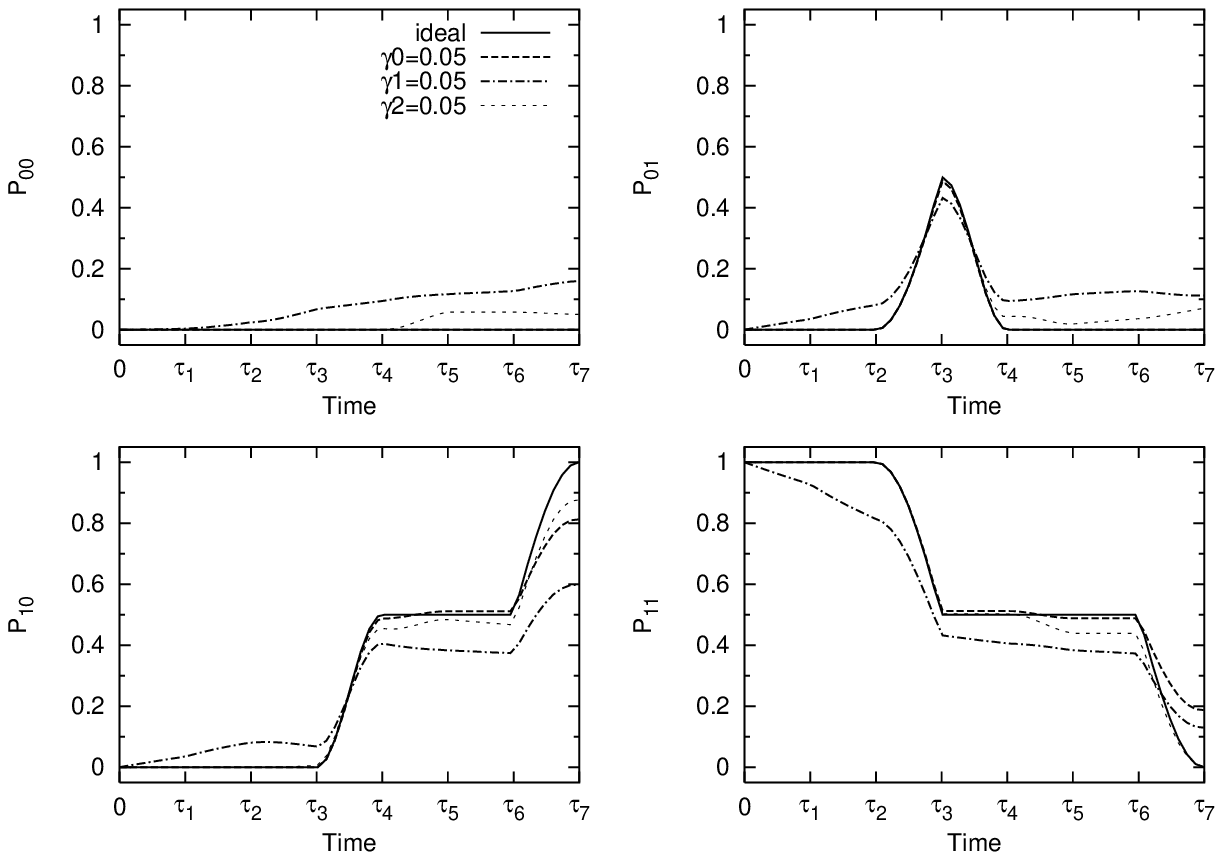}\end{center}

\caption{\label{fig:time_resolv_cnot_11} }
\end{figure}

\begin{figure}
\begin{center}\includegraphics{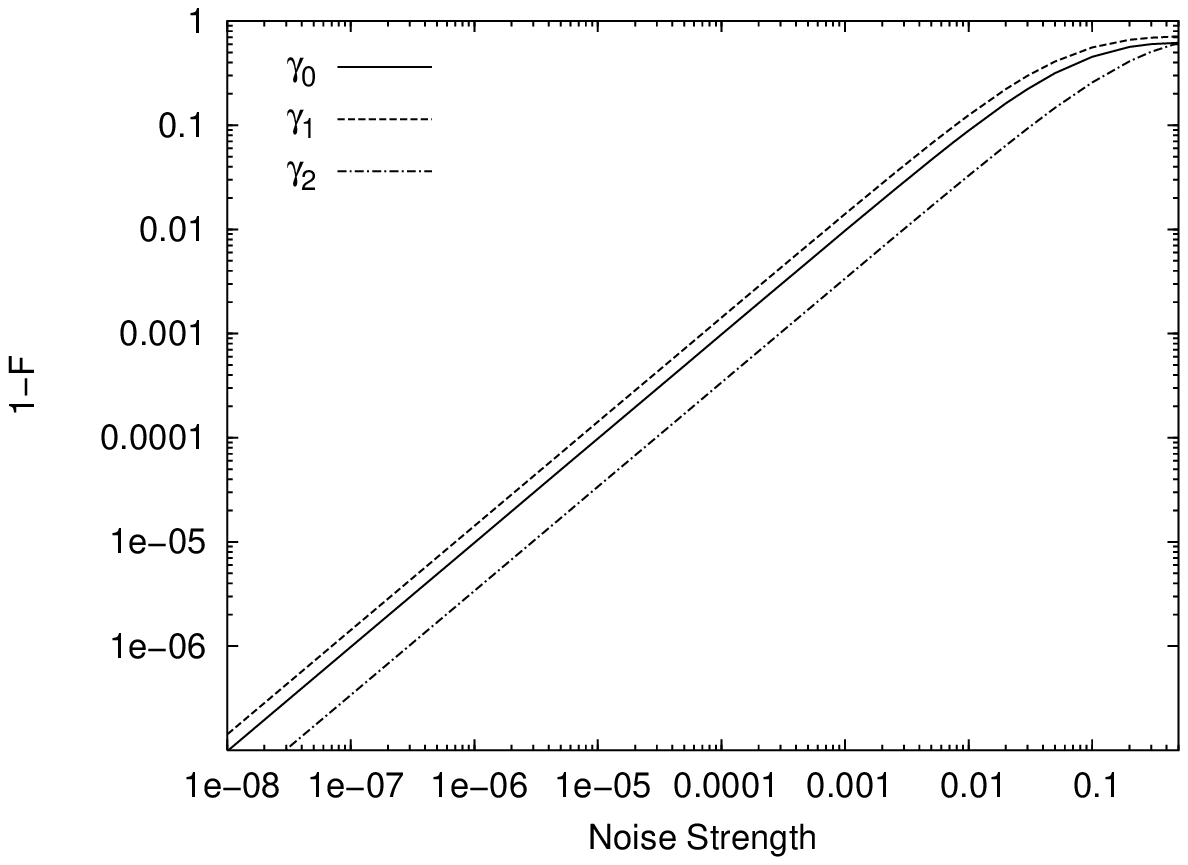}\end{center}

\begin{center}\includegraphics{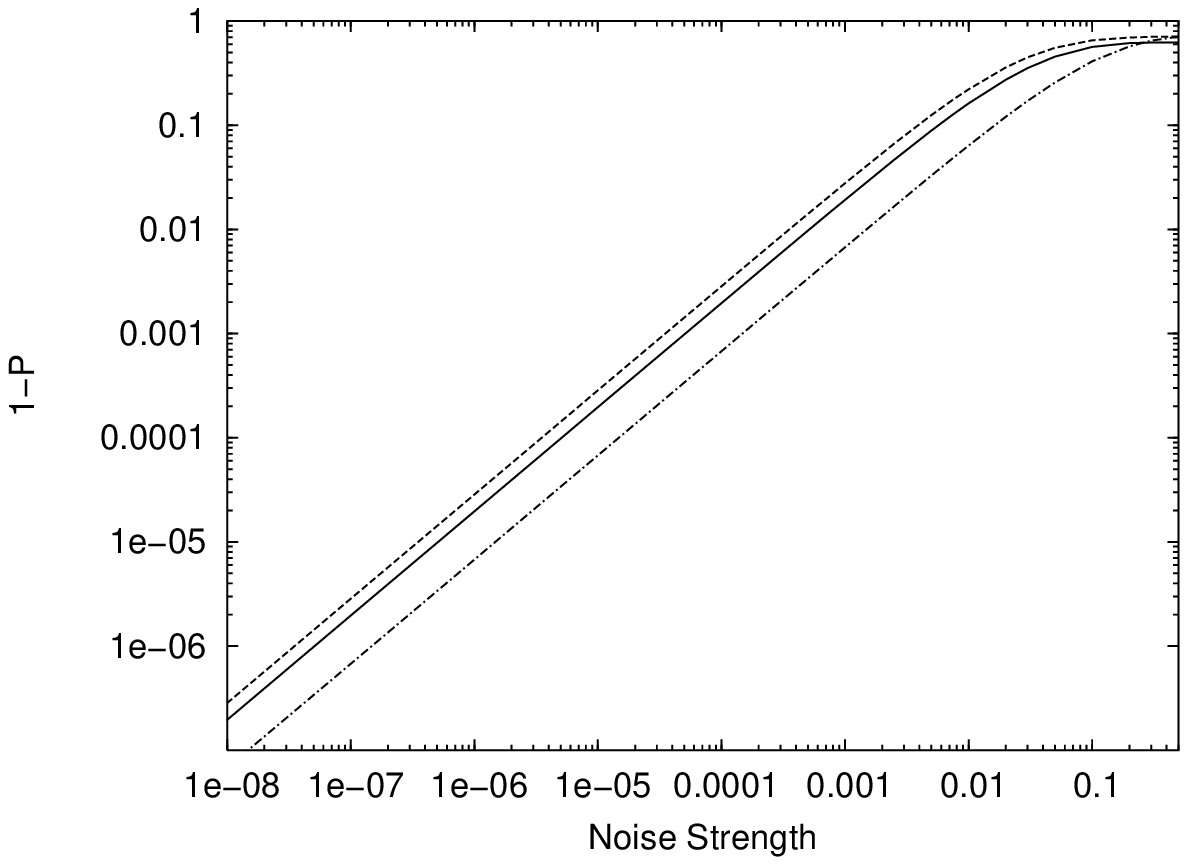}\end{center}

\caption{\label{fig:cnot_F_P} }
\end{figure}

Figure Captions:\\
\\
FIG. \ref{fig:F_tele_coherent}. Fidelity as a function of the traveling
time for the Bell states in the coherent regime: $\varepsilon_{a}=\varepsilon_{b}=1$,
$J_{a}=J_{b}=0.5$, $\gamma_{0}^{a}=\gamma_{0}^{b}=0.1$, and $\gamma_{1}^{a}=\gamma_{1}^{b}=0.1$.\\
\\
FIG. \ref{fig:F_tele_overdamp}. Fidelity as a function of the traveling
time for the Bell states in the over-damped regime: $\varepsilon_{a}=\varepsilon_{b}=0.1$,
$J_{a}=J_{b}=0.05$, $\gamma_{0}^{a}=\gamma_{0}^{b}=0.1$, and $\gamma_{1}^{a}=\gamma_{1}^{b}=0.1$.\\
\\
FIG. \ref{fig:time_resolv_cnot_11}. Tim-resolved CNOT gate operation
on the $|11\rangle$ input state. Shown are the populations in the
four basis states $\mathbf{P}_{ij}(t)=\langle ij|\rho(t)|ij\rangle$
as a function of time. The strengths of all the fields are set to
$1$ in the calculation, i.e. $\varepsilon_{0}=J_{0}=g_{0}=1$, and
the corresponding time steps are defined in Table (\ref{tab:cnot_gate_sequence}).
We show the results for four different CNOT gate operations: (1) ideal
operation without any noise (solid line), (2) operation with the strength
of the diagonal fluctuations $\gamma_{0}=0.05$ (dashed line), (3)
operation with the strength of the off-diagonal fluctuations $\gamma_{1}=0.05$
(dash-dotted line), (4) operation with the strength of the inter-qubit
coupling fluctuations $\gamma_{2}=0.05$ (dotted line).\\
\\
FIG. \ref{fig:cnot_F_P}. Dependence of the errors in the CNOT gate
operation on the noise strength. The deviations of the gate fidelity
(upper panel) and gate purity (lower panel) from the ideal values
are shown, i.e. $1-F$ and $1-P$. The effects of three types of noise
are shown in both plots: (1) diagonal fluctuations represented by
$\gamma_{0}$ (solid line), (2) off-diagonal fluctuations represented
by $\gamma_{1}$ (dashed line), (3) inter-qubit fluctuations represented
by $\gamma_{2}$ (dash-dotted line). The control-field strengths are
set to $\varepsilon_{0}=J_{0}=g_{0}=1$. \\
\\
\newpage 

\begin{figure}
\begin{center}\includegraphics{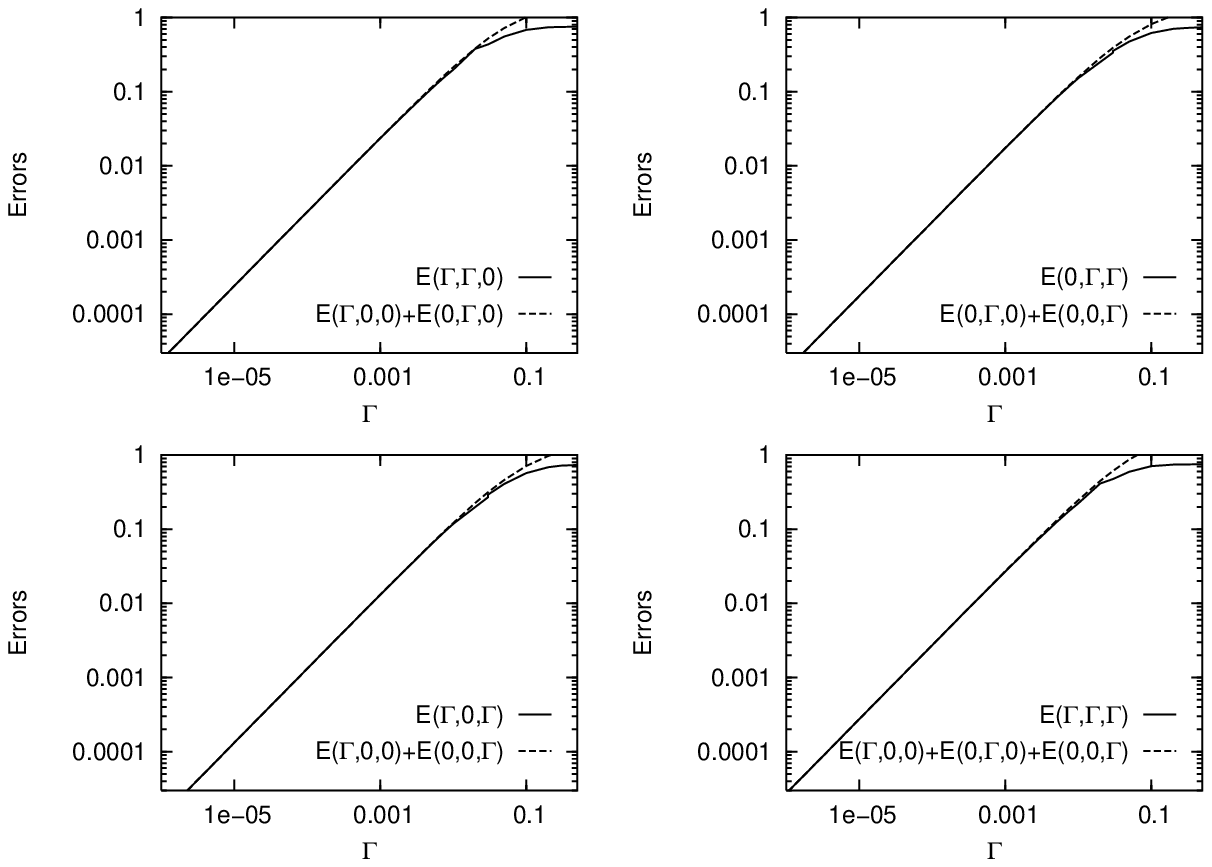}\end{center}

\caption{\label{fig:cnot_additivity} }
\end{figure}

\begin{figure}
\begin{center}\includegraphics{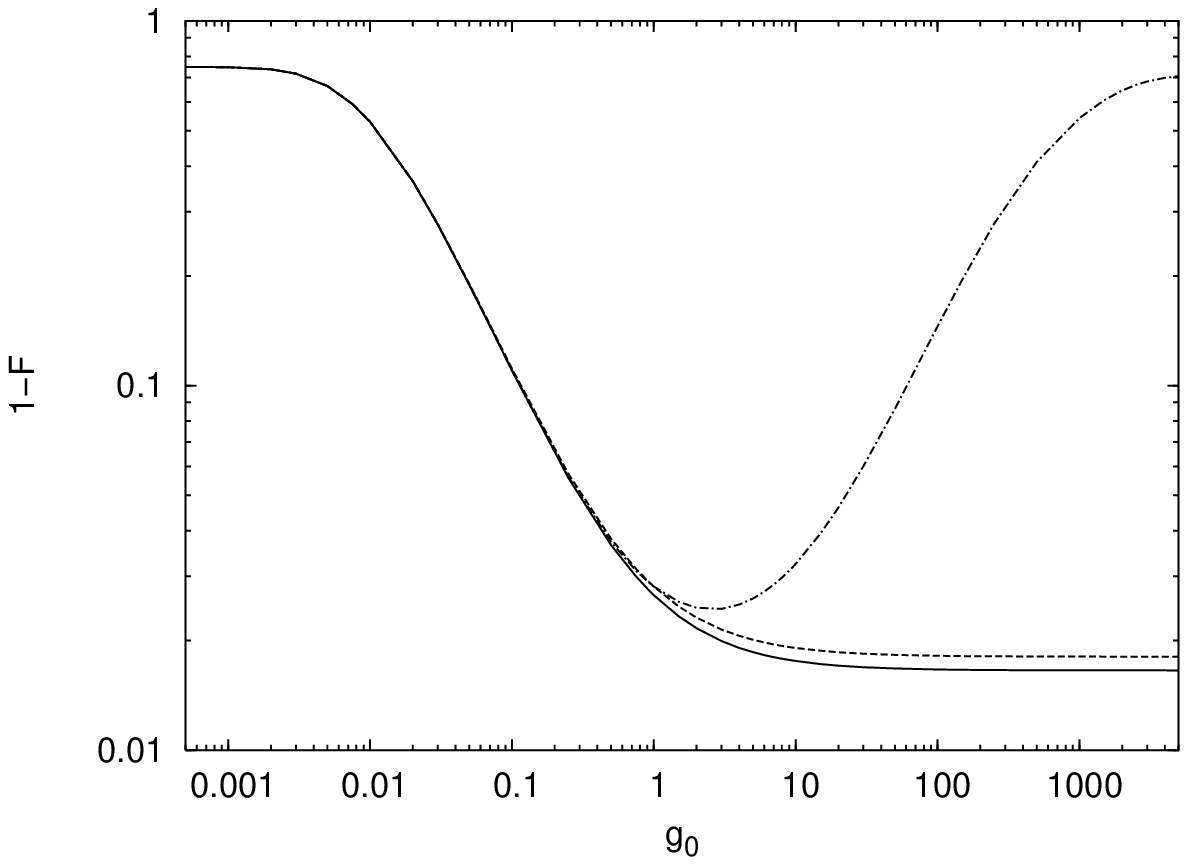}\end{center}

\caption{\label{fig:cnot_g_dependence} }
\end{figure}

~~\\
FIG. \ref{fig:cnot_additivity}. We show the error functions $E(\gamma_{0},\gamma_{1},\gamma_{2})$
of the CNOT gate operation in situations where the different types
of noise coexist (solid lines). For each case, the corresponding total
errors obtained by adding up the errors caused by the individual types
of noise is also shown (dotted lines). Four different combinations
are compared: upper-left: $E(\Gamma,\Gamma,0)$ \emph{vs.} $E(\Gamma,0,0)$+$E(0,\Gamma,0)$
($\gamma_{0}$ and $\gamma_{1}$); upper-right: $E(0,\Gamma,\Gamma)$
\emph{vs.} $E(0,\Gamma,0)$+$E(0,0,\Gamma)$ ($\gamma_{1}$ and $\gamma_{2}$);
lower-left: $E(\Gamma,0,\Gamma)$ \emph{vs.} $E(\Gamma,0,0)$+$E(0,0,\Gamma)$
($\gamma_{0}$ and $\gamma_{2}$); lower-right: $E(\Gamma,\Gamma,\Gamma)$
\emph{vs.} $E(\Gamma,0,0)$+$E(0,\Gamma,0)$+$E(0,0,\Gamma)$ (all
types of noise). The strengths of all the control fields are set to
$1$, i.e. $\varepsilon_{0}=J_{0}=g_{0}=1$. We can clearly see that
errors caused by different types of noise are additive in the small
noise regime.\\
\\
FIG. \ref{fig:cnot_g_dependence}. Dependence of the errors in the
CNOT gate operation on the strength of the inter-qubit coupling $g_{0}$.
Shown are the deviations of the gate fidelity from the ideal value
for three types of $\gamma_{2}$: (i) constant $\gamma_{2}=0.001$
(solid curve), (ii) linear $\gamma_{2}=0.001\cdot(1+g_{0})$ (dashed
curve), and (iii) quadratic $\gamma_{2}=0.001\cdot(1+g_{0}^{2})$
(dash-dotted curve). Other parameters are set to $\gamma_{0}=0.001$,
$\gamma_{1}=0.001$, $\varepsilon_{0}=1$, and $J_{0}=1$.

\clearpage \newpage

\clearpage \newpage

\clearpage \newpage

\clearpage \newpage

\clearpage \newpage

\clearpage \newpage

\end{document}